\documentclass[10pt,aps,pra,twocolumn,floatfix,showpacs,preprintnumbers,square,numbers,nofootinbib]{revtex4-2}

\usepackage[caption=false]{subfig}
\usepackage{graphicx}  
\usepackage{multirow}

\usepackage{fancyhdr}
\usepackage{parskip}
\usepackage{pgfplots}
\usepackage[T1]{fontenc}
\usepackage{dcolumn}   

\usepackage{array}
\usepackage{braket}
\usepackage{bbold}
\usepackage{booktabs}
\usepackage{tabularx}

\newcolumntype{C}{>{\centering\arraybackslash}X}      
\newcolumntype{N}{>{\centering\arraybackslash}p{1.5cm}} 

\usepackage[version=4]{mhchem}
\usepackage{standalone}
\usepackage{tikz}
\usepackage{hyperref}
\usepackage[capitalise]{cleveref}
\usetikzlibrary{decorations.pathreplacing}
\usetikzlibrary{calc}

\pgfplotsset{compat=1.17}

\newcommand{\pwisein}{\left\{ \begin{array}{ll}}
\newcommand{\pwiseout}{\end{array}\right.}

\begin{document}
\title{Mechanisms of Anomalous Three-Body Loss in a Population-Imbalanced Three-Component Fermi Gas}
\author{Kajsa-My Tempest and Chris H. Greene}
\affiliation{Department of Physics and Astronomy, Purdue University}
\date{\today}                     

\begin{abstract}  
Achieving precise control of ultracold atomic gases requires a detailed understanding of atom loss mechanisms. Motivated by the anomalous three-body decay in a three-component Fermi gas reported in Ref.~\cite{Schumacher2023}, this work investigates mechanisms that possibly contribute to the observed loss. The three-body Schrödinger equation is solved in the hyperspherical adiabatic representation with pairwise van der Waals interactions, and the $S$-matrix is obtained via the eigenchannel $R$-matrix method to compute recombination rate coefficients $K_3$ and two-body cross sections. At the magnetic field strength where the anomalous decay occurs, $K_3$ is unitary limited, exhibiting the threshold energy scaling $K_3(E)\propto E^{-1}$ that applies when only one scattering length is resonant. Consequently, the thermally averaged $\langle K_3 \rangle$ acquires a temperature dependence. Because the experiment is performed in the degenerate regime, $\langle K_3 \rangle$ also explicitly depends on the per-spin densities through the per-spin Fermi energies $E_{F}^{(i)}\propto n_i^{2/3}$. As the gas is diluted and degeneracy is reduced, $\langle K_3 \rangle$ approaches the non-degenerate value and becomes a function of temperature only. Channel-resolved branching ratios and cross sections are folded into a Monte Carlo cascade simulation of secondary collisions and trap escape. The analysis indicates that typical three-body recombination events remove fewer than three atoms on average, and that the atom losses are primarily due to the ejection of secondary collision products, rather than the initial three-body recombination products. Therefore, a significant fraction of the released binding energy remains in the trapped ensemble as kinetic energy. Retained energy drives evaporative loss, offering a plausible, partial explanation for the anomalous decay.
\end{abstract}


\maketitle 

\section{Introduction}
Ultracold atomic clouds offer a unique medium for exploring the fundamental behavior of nature at the smallest scales, enabling direct investigation of quantum phenomena at the atomic level. In this regime, atoms behave as quantum objects, and their interactions can be precisely controlled using magnetic or optical Feshbach resonances. For instance, in an external magnetic field, the interactions of bosons and two-component fermions, i.e., fermions with two different hyperfine spins, are dominated by $s$-wave interactions. The $s$-wave scattering length is then a function of the magnetic field strength \cite{Moerdijk1995}
\begin{equation}
    a(B)=a_{\rm bg}\bigg(1-\frac{\Delta}{B-B_0}\bigg),
\end{equation}
where $a_{\rm bg}$ is the background scattering length, $B_0$ is the magnetic field strength where $a$ diverges, and $\Delta $ is the resonance width. By controlling the composition of the gas, whether composed of bosons, fermions, or mixtures thereof, and tuning both interactions and density, one can create exotic states of matter, such as Bose--Einstein condensates (BECs) using bosonic atoms or novel quantum phases in two- or multi-component Fermi gases \cite{Kagan2013, Regal2007}. For example, tuning interactions in a two-component Fermi gas through a Feshbach resonance enables a smooth crossover from a Bardeen--Cooper--Schrieffer (BCS) state of weakly bound fermion pairs at negative scattering lengths ($a<0$) to a molecular Bose--Einstein condensate composed of tightly bound pairs at positive scattering lengths ($a>0$). Additionally, two- and three-component Fermi gases provide systems to investigate conditions akin to those encountered in nuclear matter and astrophysical environments \cite{Ohashi2020, Baroni2024}.

\begin{figure}[htbp]
    \centering
    \includegraphics[width=1.\linewidth]{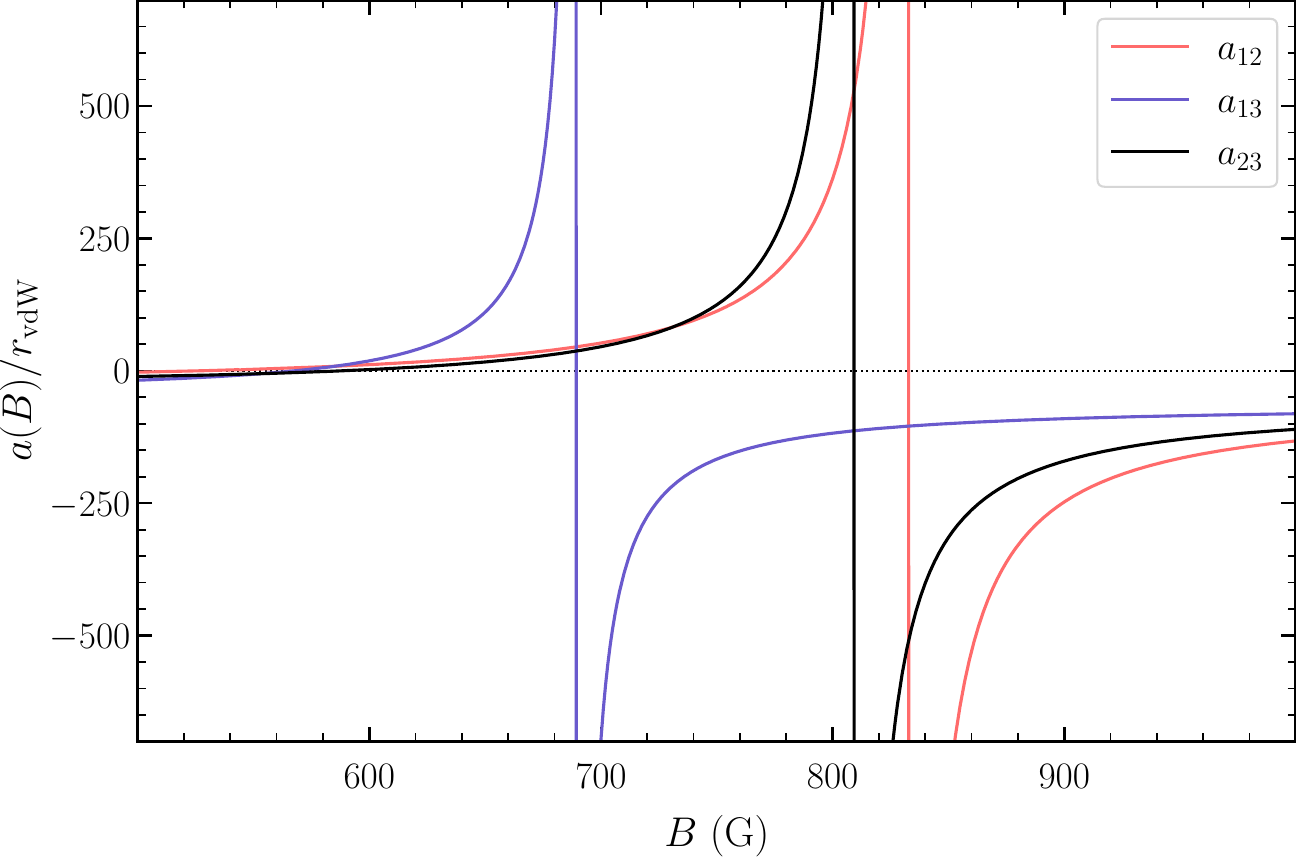}
    \caption{Scattering lengths $a(B)/r_{\rm vdW}$ for each spin pair as a function of magnetic field strength. The curves show the magnetic-field dependence of the two-body scattering lengths between the three lowest spin states of $^6$Li, expressed in units of the van der Waals length $r_{\rm vdW}$. Data adapted from Ref.~\cite{Zurn2013}.}
    \label{fig:scat}
\end{figure}

The three-component Fermi gas of $^6$Li atoms prepared in its three lowest Zeeman states offers a particularly rich system for investigating both few-body and many-body quantum phenomena. In the electronic ground state, the valence electron of a $^6$Li atom (spin $s = \textstyle \frac{1}{2}$) couples to the nuclear spin ($i = 1$), resulting in six internal hyperfine states. The three lowest states at $B=0$ correspond to $\ket{1} = \ket{\textstyle \frac{1}{2}, \textstyle \frac{1}{2}}$, $\ket{2} = \ket{\textstyle \frac{1}{2}, -\textstyle \frac{1}{2}}$, and $\ket{3} = \ket{\textstyle \frac{3}{2}, -\textstyle \frac{3}{2}}$ in the $\ket{f, m_f}$ basis, where $\vec f = \vec i + \vec s$ and $m_f$ is the projection of $\vec f$ onto the quantization axis. At magnetic fields above $B \simeq 200$ G, these states become approximately electron spin-polarized with $m_s = \textstyle -\frac{1}{2}$, and differ only in their nuclear spin projection \cite{Stoof1996, Huckans2009}. In this system, each of the three spin pairs exhibits magnetically tunable interactions through broad, overlapping Feshbach resonances, allowing access to a wide range of $s$-wave scattering length combinations. At high magnetic field strengths, above the resonances illustrated in \cref{fig:scat}, all three scattering lengths become negative and similar in magnitude, resulting in an approximate SU(3) symmetry.\footnote{Strict SU(3) symmetry requires all three two-body scattering lengths to be equal.} In this regime, several competing mechanisms emerge to reduce the energy of the many-body state: Atoms in different internal states can form trimer states in a process similar to baryon formation \cite{Rapp2007} or form Cooper pairs. The latter can occur symmetrically, leading to a color superfluid (CSF), or asymmetrically, when only two components pair while the third remains a Fermi liquid \cite{Rapp2007, Ozawa2010, Tajima2019}. A degenerate gas in this high-field SU(3)-symmetric regime exhibits a phase diagram reminiscent of that found in quantum chromodynamics (QCD), making it a compelling analog system for studying phenomena relevant to dense quark matter \cite{OHara2011, Kurkcuoglu2018, Tajima2019}. Population imbalances play a crucial role in this context, since they influence the competition between pairing channels, induce spontaneous symmetry breaking, and give rise to exotic quantum phases. In particular, theoretical studies have shown that population imbalance can arise spontaneously in SU(3)-symmetric systems due to the interplay of competing pairing mechanisms \cite{Rapp2007, Ozawa2010, Salasnich2011}. At intermediate magnetic fields, the system passes through three BCS-BEC crossovers, one for each spin pair. Below all three resonances, where the scattering lengths become positive, all three spin pairs can bind into shallow Feshbach molecules, allowing for studies of molecular Bose--Einstein condensation in multicomponent systems.

Of relevance for all experimental studies involving quantum gases is the stability of the system. The dominant loss mechanism in these systems is typically three-body recombination, in which three atoms collide and form a bound dimer and a free atom. If the recombination products carry kinetic energy exceeding the trap depth, they escape, leading to atom loss. In two-component Fermi gases, such losses are strongly suppressed due to the Pauli exclusion principle, which reduces the likelihood of three fermions occupying the same small volume, since at least two of them are identical. As a result, these systems exhibit enhanced stability not only against three-body recombination but also against atom-dimer and dimer-dimer inelastic collisions \cite{Petrov2003, Petrov2004}. In contrast, a three-component Fermi gas does not exhibit this protection, since all three spin states are distinguishable. This permits three-body processes to occur more readily and results in reduced stability. 

Moreover, the existence of Efimov states \cite{Efimov1970} is known to significantly affect the stability of bosonic gases by enhancing three-body recombination at magnetic fields where $a$ is negative and a trimer state crosses the three-atom continuum threshold. At positive $a$, Efimov trimers can lead to interference minima in three-body recombination \cite{Kraemer2006}, as well as a resonant enhancement of the atom-dimer relaxation rate when a trimer state couples to the atom-dimer continuum \cite{Knoop2009, Lompe2010}. Despite Efimov states being absent in two-component Fermi systems due to Pauli suppression\footnote{While a two-component fermionic three-body system can support two resonant $s$-wave interactions between distinguishable spin states, identical fermions interact only via higher partial waves such as $p$-waves, which are strongly suppressed at ultracold temperatures due to the centrifugal barrier.}\cite{Chen2024} they can exist in three-component Fermi gases, where all spin states are distinguishable. Experimental evidence for the ground and first excited Efimov states has been reported in a three-component Fermi gas of $^6$Li \cite{Lompe2010, Lompe2010b}.

Although atom loss due to three-body recombination in
three-component Fermi gases of $^6$Li has been extensively studied both experimentally \cite{Ottenstein2008, Huckans2009, Lompe2010, Lompe2010b, Huang2014} and theoretically \cite{Blume2008, Naidon2009, Braaten2009, Braaten2010, Rittenhouse2010, Naidon2011}, the system continues to show unexpected behavior. A recent experiment \cite{Schumacher2023} revealed an anomalous decay pattern in an imbalanced three-component Fermi gas, with strikingly different loss dynamics at two magnetic field strengths. At the higher field ($B = 845$~G), the decay was well described by the conventional three-body recombination model for both balanced and imbalanced mixtures,
\begin{equation}\label{eq:standard_rate_eq}
\frac{dn_i}{dt}=-L_3 n_1 n_2 n_3,
\end{equation}
where $L_3$ is the experimentally observed three-body recombination atom-loss coefficient and $n_i$ is the number density of atoms in spin state $i$. In contrast, near unitarity for the spin states $\ket{1}$–$\ket{3}$ ($B = 690$~G), the decay deviated strongly from standard rate-equation predictions when the initial populations were imbalanced. To reproduce the observed dynamics, an additional term proportional to the relative population of each species was introduced by Ref.~\cite{Schumacher2023}:
\begin{equation}\label{eq:phenomenological_rate_eq}
\frac{dn_i}{dt}
= -3 L_3\left(\eta\frac{n_i}{n_{\rm tot}} + \frac{1-\eta}{3}\right)n_1 n_2 n_3,
\end{equation}
where the parameter $\eta<1$ showed a density dependence and decreased with decreasing total density $n_{\rm tot}=n_1+n_2+n_3$. This modification means that the loss rate of each component scales with its relative population. Such behavior points to physics beyond simple three-body recombination, potentially involving additional few-body or many-body mechanisms, and provides the central motivation for the present theoretical investigation.

\section{Ultracold Chemical Reactions}
In an ultracold system consisting only of atoms, the dominant loss channel is three-body recombination involving all three spin states, leading to the formation of either deeply bound molecules
\begin{equation}
    \label{re:1a}
    \ce{ \ket{1} + \ket{2} + \ket{3} ->[{$K_3^{\mathrm{d}}$}] \ket{ij}_{\mathrm{d}} + \ket{k}} + \varepsilon_{\mathrm{d}}^{(ij)},
\end{equation}
or shallow Feshbach molecules:
\begin{equation}
    \label{re:1b}
    \ce{ \ket{1} + \ket{2} + \ket{3} <=>>[{$K_3^{(ij)}$}][{$D_{ij}$}] \ket{ij}_{\mathrm{s}} + \ket{k}} + \varepsilon_{\mathrm{s}}^{(ij)}.
\end{equation}

Here, $\varepsilon_{\mathrm{d}}^{(ij)}$ and $\varepsilon_{\mathrm{s}}^{(ij)}$ denote binding energies, the $K_3$ coefficients are three-body recombination rate coefficients, and $D_{ij}$ are dissociation rate coefficients. The rate coefficient $K_3^{\mathrm{d}}$ corresponds to recombination into deeply bound molecular states, while $K_3^{(ij)}$ is associated with recombination into weakly bound Feshbach molecules involving atom pairs $(ij)$. The forward recombination process in \cref{re:1b} is allowed when $a_{ij} > 0$, while the reverse reaction can occur only if the resulting shallow dimer remains trapped, and this process is typically slow \cite{Jochim2003}. The reaction rates associated with these processes are given by
\begin{equation}
    \frac{dn_i}{dt} = -\left(K_3^{\mathrm{d}} + \sum_{i<j}^3 K_3^{(ij)}\right)n_1 n_2 n_3,
\end{equation}
and
\begin{equation}
    \frac{dn_{ij}}{dt} = K_3^{(ij)} n_1 n_2 n_3 - D_{ij} n_{ij} n_{\mathrm{tot}},
\end{equation}
where $n_{ij}$ is the number density of shallow dimers formed from states $i$ and $j$. For $a_{ij} < 0$, the corresponding recombination rate vanishes, i.e., $K_3^{(ij)} = 0$.

At magnetic field strengths where all three scattering lengths are negative, three-body recombination can lead only to the formation of deeply bound dimers; that is, \cref{re:1a} is the only available pathway. The energy released in this process is large, resulting in the loss of both recombination products from the trap. Indeed, in the experiment reported in \cite{Schumacher2023}, the decay of the atomic populations at the higher field $B = 845$~G, where all three scattering lengths are negative (see \cref{fig:scat}), was consistent with that of three-body recombination involving all spin states, followed by the loss of all participating atoms.

In contrast, at $B = 690$~G, where the anomalous decay was observed, the scattering lengths are $a_{12} = 1423.1 a_0$, $a_{23} = 1177.0 a_0$, and $a_{13} = -647677.4 a_0$ \cite{Zurn2013}, where $a_0$ is the Bohr radius. The two dominant recombination processes thus form $(12)$ and $(23)$ dimers, each accompanied by the corresponding recoiling atom. The fate of the recombination products now depends on the depth $U_{\rm trap}$ of the trap.

Assuming a negligible initial kinetic energy, the binding energy $\varepsilon_{\rm b}$ released in a recombination event is divided so that the dimer acquires the kinetic energy $\textstyle \frac{1}{3}\varepsilon_{\rm b}$ and the atom $\textstyle \frac{2}{3}\varepsilon_{\rm b}$. Moreover, in a red-detuned optical dipole trap, the trapping potential is proportional to the polarizability. Because the polarizability of a dimer is approximately the sum of the polarizabilities of its constituent atoms, the dimer will experience a trapping potential that is twice as deep as that of the unbound atoms. Thus, if the binding energy satisfies $\varepsilon_{\rm b} < 6 U_{\rm trap}$, the dimer will remain trapped, and if $\varepsilon_{\rm b} < \textstyle \frac{3}{2} U_{\rm trap}$, both the atom and the dimer will remain trapped. However, in experimental settings, these conditions may not hold precisely and, for the trapping potential used in \cite{Schumacher2023}, dimers associated with $a > 1300 a_0$ (corresponding to $\varepsilon_{\rm b} \approx 10 U_{\rm trap}$) have been shown to remain trapped to some extent \cite{Ji2022}.

If a dimer does not immediately escape the trap, it can participate in secondary collisions, such as atom-dimer relaxation. Even when the dimer possesses sufficient kinetic energy to leave, additional interactions may occur if the atom-dimer cross section is large. In particular, if the mean free path is shorter than the distance to the trap edge, the dimer may scatter before escaping. Elastic collisions with atoms can decelerate the dimer and lead to the ejection of additional atoms, while inelastic collisions can relax the dimer into a more deeply bound molecular state, potentially resulting in the loss of all recombination products. These scattering processes, elastic, inelastic, and reactive, are formally described by the scattering matrix $S$, which encodes all information about a scattering event by relating the amplitudes and phases of incoming and outgoing asymptotic states.

For instance, the partial three-body recombination rate coefficient for three atoms with the total angular momentum $J$ and parity $\Pi$ is given by \cite{Suno2002}
\begin{equation}\label{eq:K3}
    K_3^{J\Pi} = \sum_{\alpha, \beta} \frac{32\hbar\pi^2 N! (2J+1)}{\mu k_{\alpha}^4} \left| S^{J\Pi}_{\beta \leftarrow \alpha} \right|^2,
\end{equation}
where $N$ is the number of identical atoms, $\mu$ is the three-body reduced mass, and $k_{\alpha}$ is the incoming three-body wave number. Here, $\alpha$ and $\beta$ denote the incoming and outgoing channels, respectively. 

The collision-induced dissociation rate for a shallow dimer $(ij)$ and atom $k$ is given by
\begin{equation}\label{eq:Dij}
    D_{ij}^{J\Pi} = \sum_{\beta} \frac{(2J+1)\pi}{\mu_{k(ij)} \, k_{k(ij)}} \left| S^{J\Pi}_{\beta \leftarrow \alpha_{ij}} \right|^2,
\end{equation}
where $\mu_{k(ij)}$ is the reduced mass of the atom-dimer system and $k_{k(ij)}$ is the corresponding relative momentum. The index $\alpha_{ij}$ labels the incoming atom-dimer channel, while $\beta$ runs over all energetically accessible outgoing channels.

More generally, the cross section for two-body scattering between an atom and a dimer, from initial channel $\alpha$ to final channel $\beta$, is given by 
\begin{equation}\label{eq:cross section} 
\sigma^{J\Pi}_{\beta \leftarrow \alpha} = \frac{(2J+1)\pi}{k_{\alpha}^2} \left| S^{J\Pi}_{\beta \leftarrow \alpha} - \delta_{\beta\alpha} \right|^2, 
\end{equation} 
where $k_\alpha$ is the wave number associated with the incoming channel. This expression encompasses both elastic ($\beta = \alpha$) and inelastic ($\beta \neq \alpha$) processes. Specifically, the atom-dimer relaxation rate is given by \cite{DIncao2008}
\begin{equation}
    \beta_{k(ij)}^{J\Pi}=\sum_{\beta}\frac{(2J+1)\pi}{\mu_{k(ij)} \, k_{k(ij)}}|S_{\beta\leftarrow\alpha}^{J\Pi}|^2.
\end{equation}

Physically, each squared $S$-matrix element represents the transition probability between scattering channels. The unitarity of the $S$-matrix ensures that the total probability across all possible scattering outcomes is conserved. In practice, its matrix elements quantify the likelihood of transitions between asymptotic states, and calculating the $S$-matrix provides direct access to observable quantities such as recombination rates and cross sections in three-body collisions. In the present work, we compute the $S$-matrix using an $R$-matrix approach, where the $R$-matrix is extracted from the solutions to the three-body Schrödinger equation. The theoretical framework and computational method underlying this calculation are outlined in the following section.

\section{Modeling a Three-Component Fermi System with van der Waals Interactions}
Atom-atom interactions exhibit long-range behavior dominated at large distances by the van der Waals interaction, $v(r) \rightarrow -C_6/r^6$, where $C_6$ is the dispersion coefficient. The associated length scale is the van der Waals length, $r_{\mathrm{vdW}} = \textstyle\frac{1}{2}(2\mu_{2\rm b} C_6/\hbar^2)^{1/4}$, with the two-body reduced mass $\mu_{2\rm b}$. When tunable via a broad Feshbach resonance, such interactions can be accurately modeled using a single-channel potential comprising an attractive van der Waals tail and a repulsive core, with a tunable short-range parameter that sets the scattering length. For the three-body system of distinguishable $^6$Li atoms, which features three broad Feshbach resonances, we employ a single-channel, adiabatic hyperspherical approach formulated in the modified Smith–Whitten coordinates of Johnson \cite{Johnson1980}. A concise summary of this method follows in the next subsections.

\subsection{Two-body model}
For each pair, we employ a Lennard--Jones 6-10 model, expressed in van der Waals units as 
\begin{equation}\label{eq:twobody potential}
    v(r)=-\frac{16}{r^6}\bigg(1-\frac{C}{r^4}\bigg),
\end{equation}
where the constant $C=C_{10}/C_6$ tunes $a$. For $^6\rm Li$, $C_6=1393.39\,\rm{a.u.}$ \cite{Yan1996}. We define $\phi(r)$ as the energy-normalized, regular solution of the radial Schr\"{o}dinger equation
\begin{equation}\label{eq:radial schrodinger}
    \bigg(-\frac{\hbar^2}{2\mu_{2\rm b}}\frac{d^2}{dr^2}+v(r)\bigg)\phi(r)=E\phi(r),
\end{equation}
where $E$ is the relative scattering energy. As $r\rightarrow \infty$
\begin{equation}
    \phi(r)\xrightarrow[]{r \to \infty}\sqrt{\frac{2\mu_{2 \rm b}}{\pi \hbar^2 k}}\sin{[kr+\delta(k)]},
\end{equation}
where $k=\sqrt{2 \mu_{2 \rm b}E}/\hbar$ and $\delta(k)$ is the $s$-wave phase shift. The zero-energy scattering length is then
\begin{equation}
    a=\lim_{k\rightarrow 0}-\frac{\tan{\delta(k)}}{k}.
\end{equation}

\subsection{Hyperspherical adiabatic representation}
To formulate the three‐body problem, we begin with the Hamiltonian for three atoms of mass $m_i$, interacting only through pairwise potentials $v(r_{ij})$
\begin{equation}
\hat{H}=-\sum_{i=1}^3\frac{\hbar^2}{2m_i}\nabla^2_i + \sum_{i<j}v(|\vec{r}_i-\vec{r}_j|).
\end{equation}
Here, $|\vec{r}_i-\vec{r}_j|\equiv r_{ij}$ is the distance between the atoms $i$ and $j$. Because the interactions depend only on relative distances, we can factor out the center‐of‐mass motion and describe the internal dynamics in terms of mass‐scaled Jacobi coordinates
\begin{align}
    \vec{\rho}_{ij}&=d_k^{-1}(\vec{r}_i-\vec{r}_j)\\ \nonumber
    \vec{\rho}_{k}&=d_k[\vec{r}_k-(\vec{r}_i+\vec{r}_j)/2],
\end{align}
with
\begin{equation}
    d_k^2=\frac{m_k(m_i+m_j)}{\mu M}, \,\,\, M=\sum_im_i,
\end{equation}
and $\mu^2=m_im_jm_k/M$.

By combining the two mass‐weighted Jacobi vectors into a single six‐dimensional vector, the three‐body configuration can be compactly parameterized using hyperspherical coordinates. These coordinates comprise a hyperradius $\rho=\sqrt{\rho^2_{ij}+\rho^2_{k}}$, which describes the overall size of the three-body system, and five hyperangles $\Omega$: two internal angles, $\theta$, which determines the shape of the triangle formed by the three atoms, and $\varphi$, which describes particle permutations at the triangle's vertices, plus three Euler angles $\alpha,\beta,\gamma$ corresponding to spatial rotations. Since we focus exclusively on $J=0$ states, only the internal hyperangles need to be explicitly considered. Our setup closely follows \cite{Tempest2023}, but with two adjustments: each pairwise interaction is assigned a distinct scattering length, and no symmetry constraints are imposed on the hyperangle that governs particle permutations.

After a rescaling of the wave function, $\Psi_n(\rho,\Omega)=\rho^{-5/2}\psi_n(\rho,\Omega)$, the three-body Schr\"{o}dinger equation for the internal motion is given by 
\begin{align}\label{eq:scrodinger} &\Bigg[ -\frac{\hbar^2}{2\mu}\frac{\partial^2}{\partial\rho^2}+\frac{\hbar^2}{2\mu \rho^2}\bigg(\hat{\Lambda}^2+\frac{15}{4}\bigg)+\sum_{i<j}v(r_{ij})\Bigg]\psi_n(\rho,\Omega)\\ \nonumber
&=E_n\psi_n(\rho,\Omega). 
\end{align} 
Here, $\hat{\Lambda}$ is the generalized grand angular momentum operator. In Johnson coordinates, $\hat{\Lambda}^2$ takes the form 
\begin{equation} \hat{\Lambda}^2=-\frac{4}{\sin{2\theta}}\frac{\partial}{\partial\theta}\bigg(\sin{2\theta}\frac{\partial}{\partial\theta}\bigg)-\frac{4}{\sin^2{\theta}}\frac{\partial^2}{\partial\varphi^2}. 
\end{equation}

We proceed by adopting the adiabatic representation, in which the total wave function is expanded onto a complete set of orthonormal channel functions $\Phi_{\nu}(\rho;\Omega)$, with the radial wave functions $F_{\nu n}(\rho)$ as expansion coefficients: 
\begin{equation}\label{eq:wfn} \psi_n(\rho,\Omega)=\sum_{\nu=0}^{\infty} \Phi_{\nu}(\rho;\Omega)F_{\nu n}(\rho). 
\end{equation}
The channel functions are solutions to the adiabatic equation 
\begin{equation}\label{eq:adiabatic}
\Bigg[\frac{\hbar^2}{2\mu \rho^2}\bigg(\hat{\Lambda}^2+\frac{15}{4}\bigg)+\sum_{i<j}v(r_{ij})\Bigg]\Phi_{\nu}(\rho;\Omega)=U_{\nu}(\rho)\Phi_{\nu}(\rho;\Omega), 
\end{equation} 
and the eigenvalues $U_{\nu}(\rho)$ are the corresponding adiabatic potential curves. Inserting \cref{eq:wfn} into \cref{eq:scrodinger}, projecting out $\Phi_{\mu}(\rho;\Omega)$, and integrating over the angles results in an infinite set of coupled ordinary differential equations: 
\begin{equation}
    \begin{split}\label{eq:coupledequations}
        \Bigg[-\frac{\hbar^2}{2 \mu}\frac{\partial^2}{ \partial \rho^2} + U_{\mu}(\rho) - \frac{\hbar^2}{2\mu}Q_{\mu\mu}(\rho) \Bigg]F_{n\mu}(\rho)&\\ -\frac{\hbar^2}{2\mu}\Bigg[\sum_{\nu\neq\mu}2P_{\mu\nu}(\rho)\frac{\partial}{\partial\rho} + Q_{\mu\nu}(\rho) \Bigg]F_{n\nu}(\rho)& = E_nF_{n\mu}(\rho),
    \end{split}
\end{equation}
where $P_{\mu\nu}(\rho)$ and $Q_{\mu \nu}(\rho)$ are coupling matrix elements defined as 
\begin{equation}\label{eq:coupling}
\begin{split}
    P_{\mu\nu}(\rho) &= \Big\langle \Phi_{\mu}(\rho;\Omega)\Big|\frac{\partial}{\partial\rho}\Phi_{\nu}(\rho;\Omega)\Big\rangle,\\
    Q_{\mu\nu}(\rho) &= \Big\langle \Phi_{\mu}(\rho;\Omega) \Big| \frac{\partial^2}{\partial\rho^2}\Phi_{\nu}(\rho;\Omega)\Big\rangle,\\
    P^{2}_{\mu\nu}(\rho) &= -\Big\langle \frac{\partial}{\partial\rho}\Phi_{\mu}(\rho;\Omega) \Big| \frac{\partial}{\partial\rho}\Phi_{\nu}(\rho;\Omega)\Big\rangle.
\end{split}
\end{equation}
The coupling matrix $\bar{P}$ is antisymmetric and therefore $P_{\nu\nu}=0$. The coupling matrices in \cref{eq:coupling} are related through \begin{equation} \bar{Q}=\frac{d\bar{P}}{d\rho}+\bar{P}^2, \end{equation} where the diagonal elements are $Q_{\nu\nu}=P^2_{\nu\nu}$. In principle, the coupled-channel equation is exact, but for computational efficiency, the channel basis is truncated.

\subsection{Boundary conditions}
To enforce the correct symmetry for three distinguishable particles, we allow the internal hyperangles to span the full ranges
\begin{equation}
    \theta \in \Big[0,\tfrac{\pi}{2}\Big],
\quad
\varphi \in [0,2\pi).
\end{equation}
The boundary conditions follow from the continuity relation in Johnson's hyperspherical coordinates:
\begin{equation}\label{eq:continuity}
\Phi_{\nu}(\rho;\theta,\varphi,\alpha,\beta,\gamma)
=\Phi_{\nu}\bigl(\rho;\theta,\varphi+2\pi,\alpha,\beta,\gamma+\pi\bigr).
\end{equation}
Together with the requirement that, as $\rho \to \infty$ (vanishing interactions), the adiabatic channel functions smoothly approach the hyperangular harmonics, i.e., the eigenfunctions of the squared grand angular momentum operator:
\begin{equation}\label{eq:angular}
\hat{\Lambda}^2\,\Phi_{\nu}(\theta,\varphi)
=\lambda(\lambda+4)\,\Phi_{\nu}(\theta,\varphi),
\end{equation}
with eigenfuctions $\Phi_{\nu}(\theta,\varphi)=g_{lm}(\theta)\,e^{im\varphi}$ and eigenvalues $\lambda = 2(l+m)$, where $l=l_{\vec{\rho}_{ij}}=l_{\vec{\rho}_{k}}$.  

Under a full $2\pi$ shift in $\varphi$ (or equivalently, a $\pi$ rotation about the body-fixed $z$-axis) the two Jacobi vectors invert \cite{Tempest2023}. From \cref{eq:continuity}, one then finds that for even parity
\begin{equation}
    m\in\mathbb{Z},
\quad \text{and} \quad
\Phi_{\nu}(\rho;\theta,\varphi)
=\Phi_{\nu}(\rho;\theta,\varphi+2\pi).
\end{equation}
Thus, the spectrum of eigenvalues $\lambda=0_{(1)},2_{(2)},4_{(3)},6_{(4)},8_{(5)} \ldots$, where the subscript in parentheses denotes the degeneracy of each level.

Finally, inserting the separable form $g_{lm}(\theta)e^{im\varphi}$ into \cref{eq:angular} and examining the limits $\theta \rightarrow 0$ and $\theta \rightarrow \textstyle \frac{\pi}{2}$, it follows that $g_{lm}(\theta) \sim \theta^m$ as $\theta \rightarrow 0$, and $g_{lm}(\theta)\rightarrow \text{const.}$ as $\theta \rightarrow \textstyle \frac{\pi}{2}$. For regular solutions with $m \geq 0$, Neumann boundary conditions are required:
\begin{equation}
    \frac{\partial \Phi_{\nu}}{\partial \theta} \Bigg\rvert_{\theta = 0, \frac{\pi}{2}} = 0.
\end{equation}

\subsection{Numerical implementation}
We represent each adiabatic channel function in a tensor‐product basis made up of B-splines of order $k$.  On the interval $\theta\in[0,\textstyle \frac{\pi}{2}]$ we employ standard (nonperiodic) splines with Neumann boundary conditions. Along the hyperangle $\varphi$, we build a $2\pi$-periodic spline basis by appending “ghost” knot copies at $\varphi\pm2\pi$ and summing their contributions so that each spline and its first $k-2$ derivatives match continuously at $\varphi=0$ and $2\pi$. Collocation at Gauss–Legendre points in every $(\theta,\varphi)$ cell then converts \cref{eq:adiabatic} into the generalized matrix eigenvalue problem

\begin{equation}
    \bar{H}_{\rm ad}(\rho)\,\mathbf{c}(\rho)
=U(\rho)\,\bar{B}\,\mathbf{c}(\rho),
\end{equation}
whose solutions $\{U_{\nu}(\rho),\mathbf{c}_{\nu}(\rho)\}$ yield the hyperangular eigenvalues and channel functions $\Phi_{\nu}(\rho;\theta,\varphi)$.

\subsection{Three-body effective hyperradial potentials}
We now turn to the three-body effective hyperradial potentials. These are defined as
\begin{equation}\label{eq:3BP_dia} 
    W_\nu(\rho) = U_\nu(\rho) - \frac{\hbar^2}{2\mu} Q_{\nu\nu}(\rho). 
\end{equation} 
In the asymptotic limit $\rho \gg |a_{ij}|$, the effective potentials reflect either three-body continuum configurations or atom-dimer channels. Continuum channels approach the form
\begin{equation}\label{eq:continuum_channel}
    W_\nu(\rho) \xrightarrow{\rho \to \infty} \hbar^2 \frac{\lambda(\lambda + 4) + 15/4}{2\mu\rho^2}, 
\end{equation} 
while atom-dimer channels behave as
\begin{equation}\label{eq:atomdimer_channel}
    W_\nu(\rho) \xrightarrow{\rho \to \infty} -\varepsilon_{\nu \ell'} + \hbar^2\frac{\ell(\ell + 1)}{2\mu\rho^2}, 
\end{equation} 
where $\varepsilon_{\nu \ell'}$ is the binding energy of the dimer associated with channel $\nu$ and internal angular momentum $\ell'$, and $\ell$ is the relative angular momentum between the dimer and the remaining atom.

\begin{figure}[htbp!]
    \centering
    \includegraphics[width=1.\linewidth]{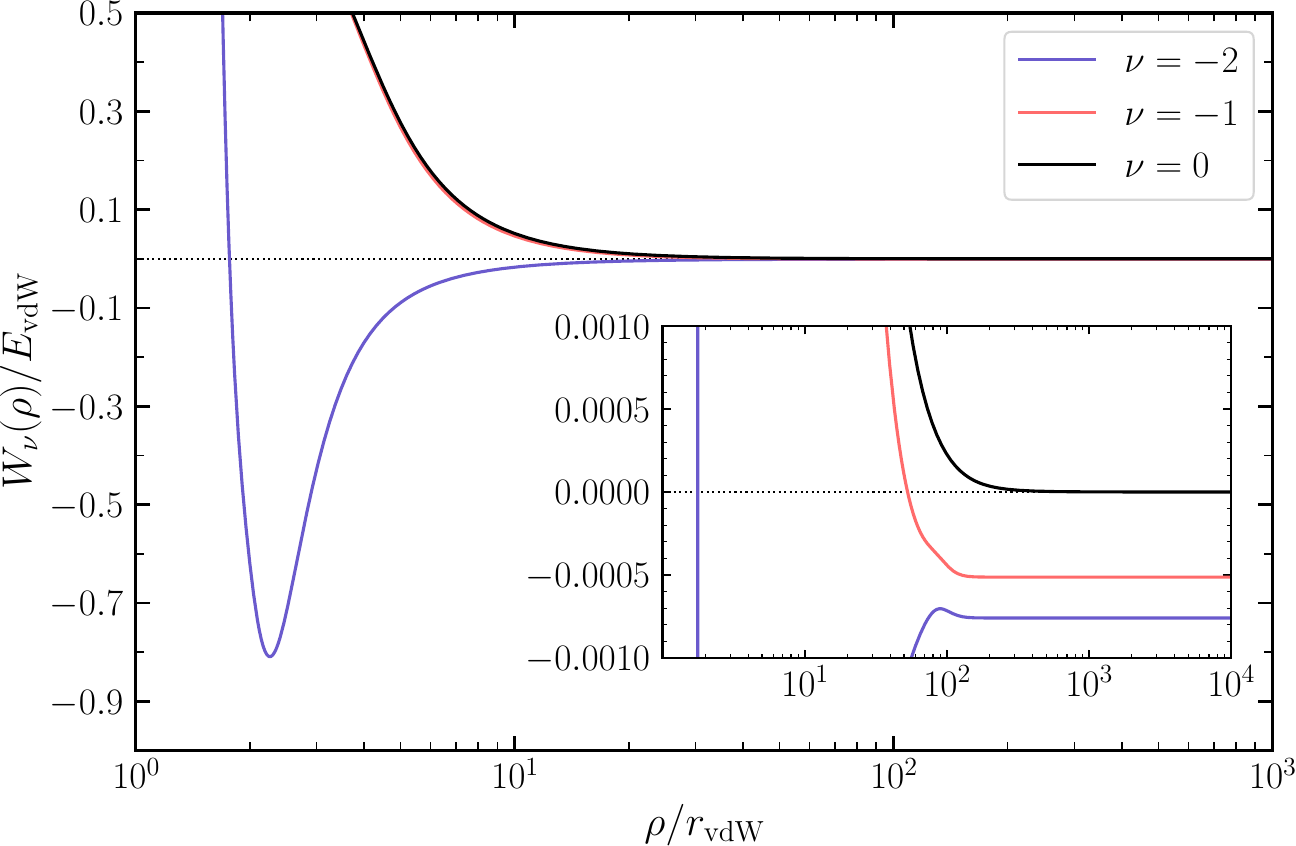}
    \caption{Effective three‐body hyperradial potentials $W_{\nu}(\rho)$ at $B = 690$~G. In the inset (zoomed-in region), the two deepest curves ($\nu=-2$ and $\nu=-1$) each asymptote at large $\rho$ to the negative binding energy of their respective Feshbach dimers, whereas the highest curve ($\nu=0$) merges into the three-atom continuum.}
    \label{fig:threebodypot}
\end{figure}

At $B = 690$ G, the effective hyperradial potentials $W_{\nu}(\rho)$ exhibit both atom-dimer and three‐atom thresholds. We classify the channels based on their asymptotic behavior: $\nu = 0$ denotes the lowest three-body continuum channel, while $\nu < 0$ identifies atom-dimer channels, with $\nu = -1$ lying closest to the dissociation threshold. As illustrated in \cref{fig:threebodypot}, the two lowest‐lying potentials ($\nu = -2$ and $\nu = -1$) asymptotically approach the negative binding energies $-\varepsilon_{-2,0}$ and $-\varepsilon_{-1,0}$ of their respective Feshbach dimers, consistent with \cref{eq:atomdimer_channel}. The uppermost branch ($\nu = 0$) remains repulsive at large $\rho$ and approaches the three‐atom continuum form given in \cref{eq:continuum_channel}.  

To analyze the behavior in the scale-free region $r_{\mathrm{vdW}} \ll \rho \ll |a_{ij}|$, we express the effective potentials in the form
\begin{equation}\label{eq:scale_free_pot}
W_\nu(\rho) = \hbar^2\frac{\xi_\nu^2 - 1/4}{2\mu\rho^2}. 
\end{equation} 
In this regime, the potentials can exhibit signatures of Efimov physics, as is known for systems of identical bosons. Here, $\xi_\nu$ is a dimensionless parameter that characterizes the strength of the effective interaction. When all three scattering lengths satisfy $|a_{ij}| \gg \rho \gg r_{\mathrm{vdW}}$, the system supports a single channel with Efimov-like attraction, characterized by $\xi_\nu = i s_0$ with $s_0 \approx 1.00624$. The value of $\xi_\nu$ depends on the mass ratios and symmetry properties of the specific three-body configuration under consideration \cite{DIncao2018}.

\begin{figure}[htbp!]
    \centering
    \includegraphics[width=1.\linewidth]{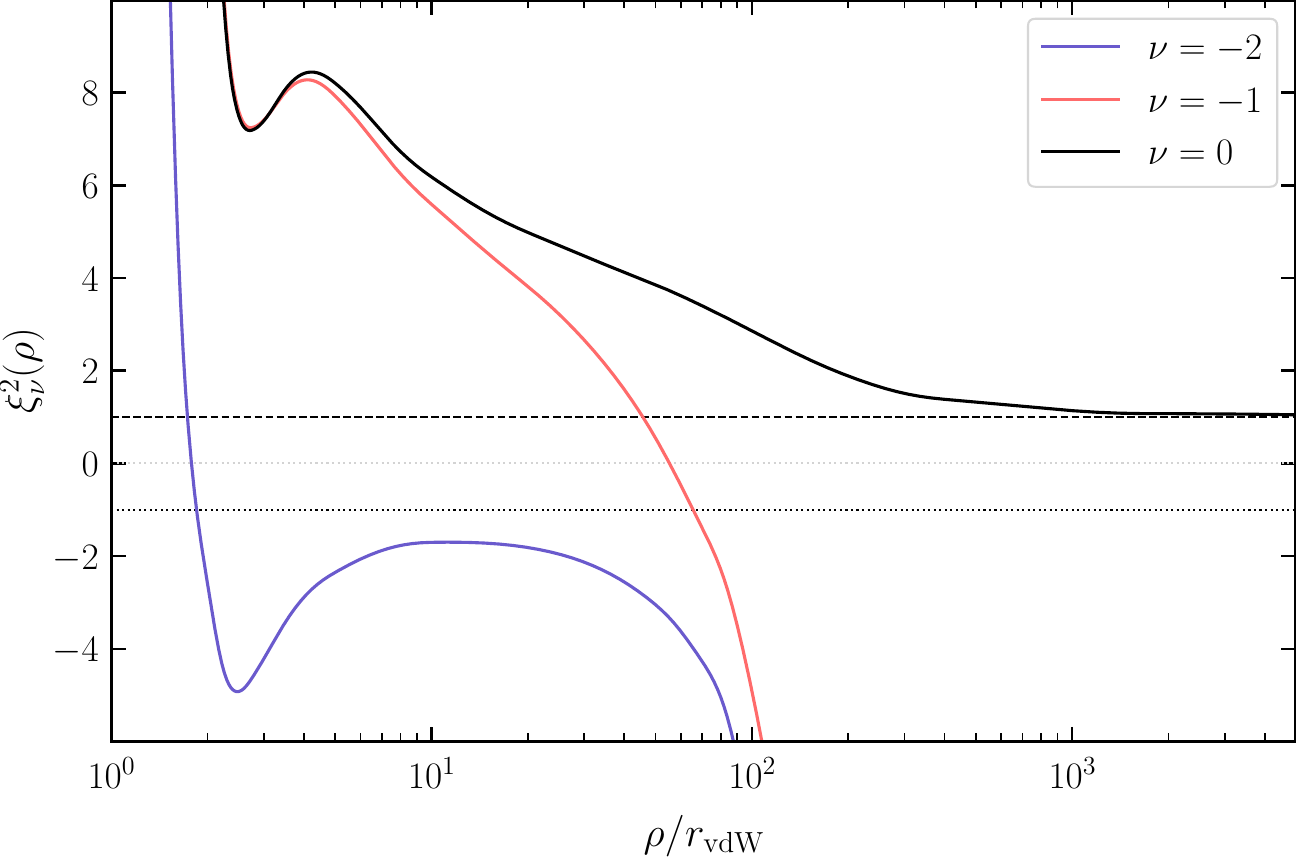}
    \caption{Rescaled effective potentials $\xi^2_{\nu}(\rho)$ computed at $B = 690$~G. The dotted black line marks the Efimovian value $-s_0^2$ for three identical bosons, while the dashed black line indicates the universal coefficient $p_0^2 = 1$ associated with Efimov-like repulsion in a three-body system of distinguishable atoms. The two upper curves have been diabatized through a peak caused by an avoided crossing in the corresponding adiabatic potential energy curves.}
    \label{fig:xi}
\end{figure}

In \cref{fig:xi}, we present the rescaled effective potentials $\xi_{\nu}^2(\rho)$, calculated using two-body interactions at $B = 690$~G. The dotted black line marks the Efimovian value $-s_0^2$ for three identical bosons, and the dashed black line indicates the universal coefficient $p_0^2 = 1$ relevant for three distinguishable atoms with $J^\Pi = 0^+$ symmetry. The $\nu = -2$ potential approximates the Efimovian form in the scale-free region $r_{\rm vdW} \ll \rho \ll a_{12}, a_{23}, |a_{13}|$, and asymptotically ($\rho \gg a_{23}$) describes a (23) dimer and a free atom in spin state 1. The $\nu = 0$ channel closely follows the repulsive Efimovian form with $\xi_0 = p_0$ in the regime $a_{12}, a_{23} \ll \rho \ll |a_{13}|$.

\begin{figure}[htbp!]
    \centering
    \includegraphics[width=1.\linewidth]{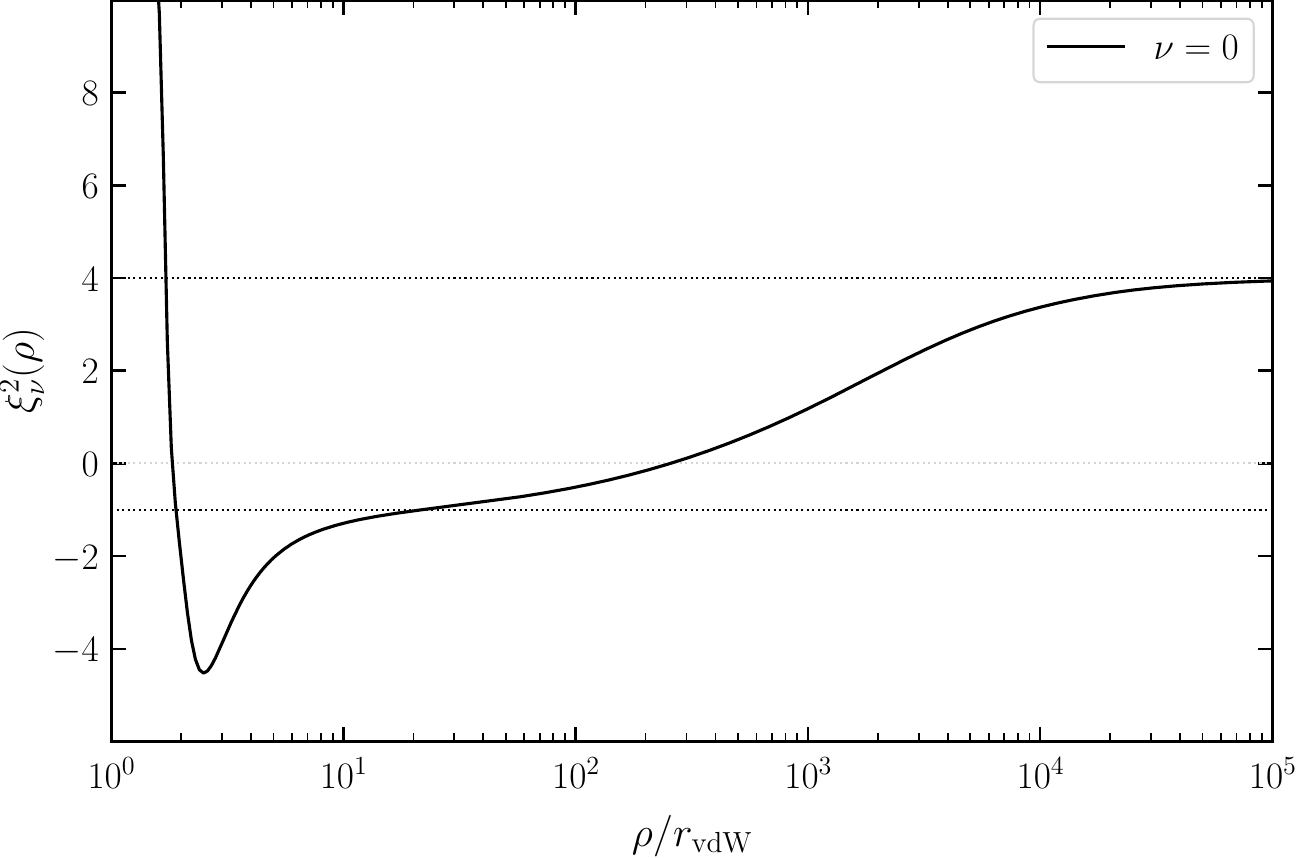}
    \caption{Rescaled effective potential $\xi_0^2(\rho)$ for the $\nu = 0$ channel at $B = 845$~G, where all pairwise scattering lengths are negative and large in magnitude. In the scale-free region $r_{\rm vdW} \ll \rho \ll |a_{ij}|$, the potential approximates the Efimovian form. At large $\rho$, it asymptotically approaches the free three-body continuum limit corresponding to $\lambda = 0$.}
    \label{fig:xi_845}
\end{figure}

In \cref{fig:xi_845}, we show the rescaled effective potential $\xi_0^2(\rho)$ for the $\nu = 0$ channel, computed using two-body interactions at $B = 845$~G, where all scattering lengths are negative and large in magnitude. This channel corresponds to the lowest three-body continuum state. In the scale-free region $r_{\rm vdW} \ll \rho \ll |a_{ij}|$, the potential approximates the Efimovian form, reflecting the effect of three near-resonant three-body interactions. At large hyperradii, it asymptotically approaches the free three-body continuum behavior characterized by $\lambda = 0$.

\subsection{\texorpdfstring{$R$-matrix}{R-matrix} calculations in the adiabatic representation}\label{sec:R-matrix}
We compute the scattering matrix using the eigenchannel $R$-matrix method \cite{Greene1983, Burke1999}, which partitions the configuration space into an internal region, where the wave function must be solved numerically, and an external region with known analytic solutions. The $R$-matrix, evaluated at a matching hyperradius $\rho_{\rm m}$, connects these regions and encodes the information needed to construct the scattering matrix.

In the eigenchannel $R$-matrix method, a special set of independent solutions to the time-independent Schr\"{o}dinger equation is obtained variationally. Each of these solutions has a constant logarithmic derivative $-b_n$ in every channel $\nu$.  The $n$th eigenchannel solution to the full $J=0$ wave function is represented in \cref{eq:wfn} for $\rho \leq \rho_m$. Moreover, in the variational calculation, each hyperradial wave function is linearly expanded in a B-spline basis set $\{ y_k \}$, i.e., $F_{\nu n}(\rho)=\sum_k y_k(\rho) c_{\nu k,n}$.

The final implementation of this basis set in the variational principle leads to a linear generalized eigenvalue problem of the form
\begin{equation}\label{eq:gamma}
    \bar\Gamma \vec{c} = b \bar{\mathcal{O}} \vec{c}.
\end{equation}
Here the matrix elements of $\bar \Gamma$ involve matrix elements of the Bloch-operator modified Hamiltonian $\hat{H}_{\rm B}$:
\begin{equation}\label{eq:gamma_elements}
    \Gamma_{\mu k', \nu k}=  \langle \Phi_\mu y_{k'}|2 \mu (E-\hat{H}_{\rm B})| \Phi_\nu y_k \rangle.
\end{equation}
If only a single radial basis function $y_k$, denoted $y_{k_0}$, is nonzero at the reaction surface $\rho_m$ and is chosen to satisfy $y_{k_0}(\rho_m)=1$, such that $y_k(\rho_m) = \delta_{kk_0}$, then the surface overlap matrix elements reduce to
\begin{equation}
    \mathcal{O}_{\mu k',\nu k}=\delta_{k'k_0}\delta_{kk_0}\delta_{\mu\nu}.
\end{equation}
As a result, the number of nontrivial eigenvalues $b_n$ is equal to the number of adiabatic channels retained in the expansion. 

When solving \cref{eq:gamma}, the radial basis functions satisfy Dirichlet boundary conditions at a small hyperradius $\rho_0 > 0$ and are defined on the finite interval $[\rho_0,\rho_{\rm m}]$. Using the Hamiltonian of \cref{eq:scrodinger} and the definitions in \cref{eq:coupling}, the variational matrix elements in \cref{eq:gamma_elements} become
\begin{equation}
    \begin{split}
        \Gamma_{\mu k', \nu k} = &\delta_{\mu\nu} \int_{\rho_0}^{\rho_{\rm m}} y_{k'}(\rho) \bigg[\hbar^2\frac{d^2}{d\rho^2} + 2\mu \Big(E - U_\nu(\rho)\Big) \bigg] y_k(\rho) \, d\rho \\
        &+ \int_{\rho_0}^{\rho_{\rm m}} \hbar^2\bigg[y_{k'}(\rho) P_{\mu\nu}(\rho) \frac{d}{d\rho} y_k(\rho) \\ 
        &- \frac{d}{d\rho}y_{k'}(\rho) P_{\mu\nu}(\rho) y_k(\rho)
        + y_{k'}(\rho) P^2_{\mu\nu}(\rho) y_k(\rho) \bigg] d\rho \\
        &- \delta_{\mu\nu} y_{k'}(\rho_{\rm m}) \hbar^2\frac{d}{d\rho} y_k(\rho_{\rm m}).
    \end{split}
\end{equation}
Further details about the implementation of the eigenchannel $R$-matrix method for this type of problem can be found in Ref. \cite{Burke1999}.

Once the eigenvalues $b$ and eigenvectors $\vec{c}$ of \cref{eq:gamma} are obtained, the $R$-matrix is constructed as
\begin{equation}
    \mathcal{R}_{\mu\nu}=-\sum_{n}b^{-1}_{n}Z_{\mu n}(Z^{-1})_{n \nu},
\end{equation}
where $Z_{\mu n}\equiv c_{\mu k_0,n}$.

At the matching hyperradius $\rho_m$, which is large enough so that we can neglect the $P$-matrix there, the radial wave functions match their asymptotic forms, expressed in terms of energy-normalized spherical Bessel and Neumann functions
\begin{equation}
    \begin{split}
    f_{\mu\nu}(\rho)&=\sqrt{\textstyle\frac{2\mu}{\pi k_\mu}} k_\mu \rho j_{l_{\mu}}(k_\mu\rho)\delta_{\mu\nu}, \\
    g_{\mu\nu}(\rho)&=\sqrt{\textstyle\frac{2\mu}{\pi k_\mu}} k_\mu \rho n_{l_{\mu}}(k_\mu\rho)\delta_{\mu\nu}
    \end{split}
\end{equation}
along with their derivatives $f'_{\mu\nu}(\rho)$ and $g'_{\mu\nu}(\rho)$. Here, $k_\mu$ and $l_\mu$ are determined by the asymptotic behavior of the corresponding effective potential. In a three-body recombination process, the continuum channels serve as incoming channels $\alpha$, with $l_\alpha = \lambda + \textstyle\frac{3}{2}$ and $k_\alpha = \sqrt{2\mu E}$, where $E$ is the total collision energy. In this case, the potential in \cref{eq:continuum_channel} has been recast into the form of a two-body centrifugal barrier with an effective angular momentum $l_\alpha$ \cite{Esry2001}. The outgoing channels $\beta$ correspond to atom-dimer recombination products, with $l_\beta = \ell$ and $k_\beta = \sqrt{2\mu(E - \varepsilon_{\beta\ell'})}$ [see \cref{eq:atomdimer_channel}], where $\varepsilon_{\beta\ell'}$ is the binding energy of the dimer in channel $\beta$. The reaction matrix is then given by
\begin{equation} 
\bar{K} = \left( \bar{f} - \bar{f}' \bar{\mathcal{R}} \right) \left( \bar{g} - \bar{g}'\bar{\mathcal{R}} \right)^{-1}, 
\end{equation} 
and the scattering matrix $\bar{S}$ is obtained via the Cayley transform:
\begin{equation} 
\bar{S} = (\mathbb{1} + i\bar{K})(\mathbb{1} - i\bar{K})^{-1}. 
\end{equation}

\section{Energy and temperature considerations}
\subsection{Threshold effects}
The threshold behavior of scattering observables in three-body systems is governed by the asymptotic form of the effective hyperradial potentials~\cite{Esry2001, Suno2002}. Threshold effects are relevant for processes that couple to the lowest three-body continuum channel, which determines the near-threshold energy dependence of observables such as recombination and dissociation rates. The transition matrix elements are defined as
\begin{equation}
    S_{\alpha\beta} = \delta_{\alpha\beta} + 2i T_{\alpha\beta},
\end{equation}
so that for recombination processes $|S_{\alpha\beta}|^2 = 4|T_{\alpha\beta}|^2$. Near threshold, the matrix elements scale as
\begin{equation}
    T_{\alpha\beta} \propto k_{\alpha}^{l_\alpha + \frac{1}{2}} k_\beta^{l_\beta + \frac{1}{2}},
\end{equation}
where $l_\alpha$ and $l_\beta$ follow from the asymptotic channel behavior discussed in \cref{sec:R-matrix}. Since $k_\beta$ remains finite, the threshold scaling is dictated solely by the incoming wave number $k_\alpha$. The recombination cross section then behaves as
\begin{equation}
    \sigma_{\beta \leftarrow \alpha}^{\rm rec} \propto \frac{|S_{\alpha\beta}|^2}{k^5} \propto k^{2\lambda - 1},
\end{equation}
and since the recombination rate coefficient scales as $K_3 \propto k \sigma_{\beta \leftarrow \alpha}^{\rm rec}$, the resulting threshold law is
\begin{equation}
    K_3 \propto k^{2\lambda} \propto E^{\lambda}.
\end{equation}
Similarly, comparison with \cref{eq:Dij} yields the threshold scaling for atom-dimer dissociation:
\begin{equation}
    D_{ij} \propto k^{2\lambda + 4} \propto E^{\lambda + 2}.
\end{equation}

In the case of three distinguishable fermions with finite non-resonant scattering lengths, the lowest continuum channel is characterized by $\lambda = 0$. However, this behavior is modified near a resonance pole. At $B = 690$~G, where $a_{13} \to \pm \infty$, the asymptotic form of the effective potential changes and is instead governed by \cref{eq:scale_free_pot}. In this regime, the effective $\lambda$ becomes 
\begin{equation} 
\lambda_{\rm eff} = -2 + \xi_0, 
\end{equation}
where $\xi_0 = 1$, see \cref{fig:xi}. This leads to recombination rate coefficients that scale like $K_3\propto E^{-1}$ and dissociation coefficients that scale like $D_{ij}\propto E$.

\subsection{Thermal averaging}
To compare with the experiment, we compute the thermally averaged three-body recombination coefficients
\begin{equation}
    \langle K_3 \rangle (T) = \int K_3(E)\, f(E)\, dE,
\end{equation}
where $f(E)$ is the collision‐energy distribution. In the classical limit $(T/T_{\rm F}\gg1)$ \cite{DIncao2004},
\begin{equation}
    f(E) = \frac{1}{2(k_{\rm B} T)^3} E^2 e^{-E/k_{\rm B} T},
\end{equation}
while in the quantum‐degenerate regime $(T/T_{\rm F}\ll1)$ the collision-energy distribution follows from \cite{Chen2022}
\begin{equation}\label{eq:Fermi-Dirac}
    \begin{split}
    f(E) =\; & Z \int \delta\left( \epsilon_1 + \epsilon_2 + \epsilon_3 - E_{\mathrm{cm}} - E \right) \\
    &\times \prod_{i=1}^{3} f_{i}(\vec{k}_i)\,d^3\vec{k}_i,
    \end{split}
\end{equation}
where $\epsilon_i = \hbar^2 k_i^2 / (2m)$ is the single-particle kinetic energy, and $E_{\mathrm{cm}} = \hbar^2 |\vec{k}_1 + \vec{k}_2 + \vec{k}_3|^2 / (6m)$ is the center-of-mass kinetic energy. The Fermi--Dirac distribution for spin component $i$ is
\begin{equation}
    f_i(\vec{k}_i) = \frac{1}{e^{(\epsilon_i - \mu_i) / k_{\rm B} T} + 1},
\end{equation}
and $Z$ is a normalization constant ensuring $\int f(E)\, dE = 1$. Each spin component $i$ has density $n_i$, Fermi wave vector $k_F^{(i)}=(6\pi^2n_i)^{1/3}$, Fermi energy $E_F^{(i)}=\hbar^2[k_F^{(i)}]^2/(2m)$, and Fermi temperature $T_F^{(i)}=E_F^{(i)}/k_B$. The finite temperature chemical potential $\mu_i$ is related to the fugacity
$z_i = e^{\mu_i/(k_BT)}$, and is determined by inverting the relation between density and fugacity,
\begin{equation}\label{eq:density_fug}
n_i\,\lambda_{\rm dB}^3
=F_{1/2}(z_i),
\end{equation}
where $\lambda_{\rm dB}=\hbar\sqrt{2\pi/(mk_{\rm B}T)}$ is the thermal de Broglie wavelength, and the half-order Fermi integral is
\begin{equation}
    F_{1/2}(z)
=\frac{2}{\sqrt\pi}\int_0^\infty\frac{\sqrt{\varepsilon}\,d\varepsilon}{z^{-1}e^{\varepsilon}+1}.
\end{equation}
Given $n_i$ and $T$, we compute $n_i\lambda_{\rm dB}^3$ and solve \eqref{eq:density_fug} for $z_i$ using a Brent rootfinder.  In the classical limit $z_i\ll1$, one recovers
$\mu_i \approx k_B T\ln\bigl(n_i\,\lambda_{\rm dB}^3\bigr)$.

We evaluate the integral in \cref{eq:Fermi-Dirac} using Monte Carlo sampling, drawing triplets of momenta from Fermi--Dirac distributions, computing their relative collision energy $E$, and averaging the energy-dependent recombination rate coefficients $K_3(E)$ over the resulting distribution.

\section{Scope and Limitations}
The loss model developed in the following section uses three-body recombination rates, both atom-atom and atom-dimer cross sections, and samples initial atomic momenta from noninteracting Fermi--Dirac distributions. In the strongly interacting and degenerate regime at $B=690$~G, many-body effects may modify both the collision dynamics and the momentum distributions. A full treatment of such medium corrections requires a many-body description and lies beyond the scope of this study. In what follows, we therefore implement only the few-body physics described above to model the decay dynamics.

\section{Results and Discussion}
This section presents our main numerical findings for the three-body system of $^6{\rm Li}$ in three different spin states. Firstly, the binding energy spectrum is mapped as a function of the magnetic field strength. Next, thermally averaged recombination rate coefficients and branching ratios, together with the elastic and inelastic two-body cross sections that govern secondary collisions, are computed. These results enable a computation of the mean free paths of recoiling atoms and emergent dimers, and an evaluation of how the magnitude and energy dependence of these cross sections shape the secondary-collision probabilities and trap-loss dynamics. Lastly, the calculated branching ratios and cross sections are used in a Monte Carlo cascade model of secondary collisions and trap escape to predict the time evolution of spin-resolved atom loss for both balanced and imbalanced initial spin-density conditions.

\subsection{Three-body binding energies}
\begin{figure}[htbp!]
    \centering
    \includegraphics[width=1.\linewidth]{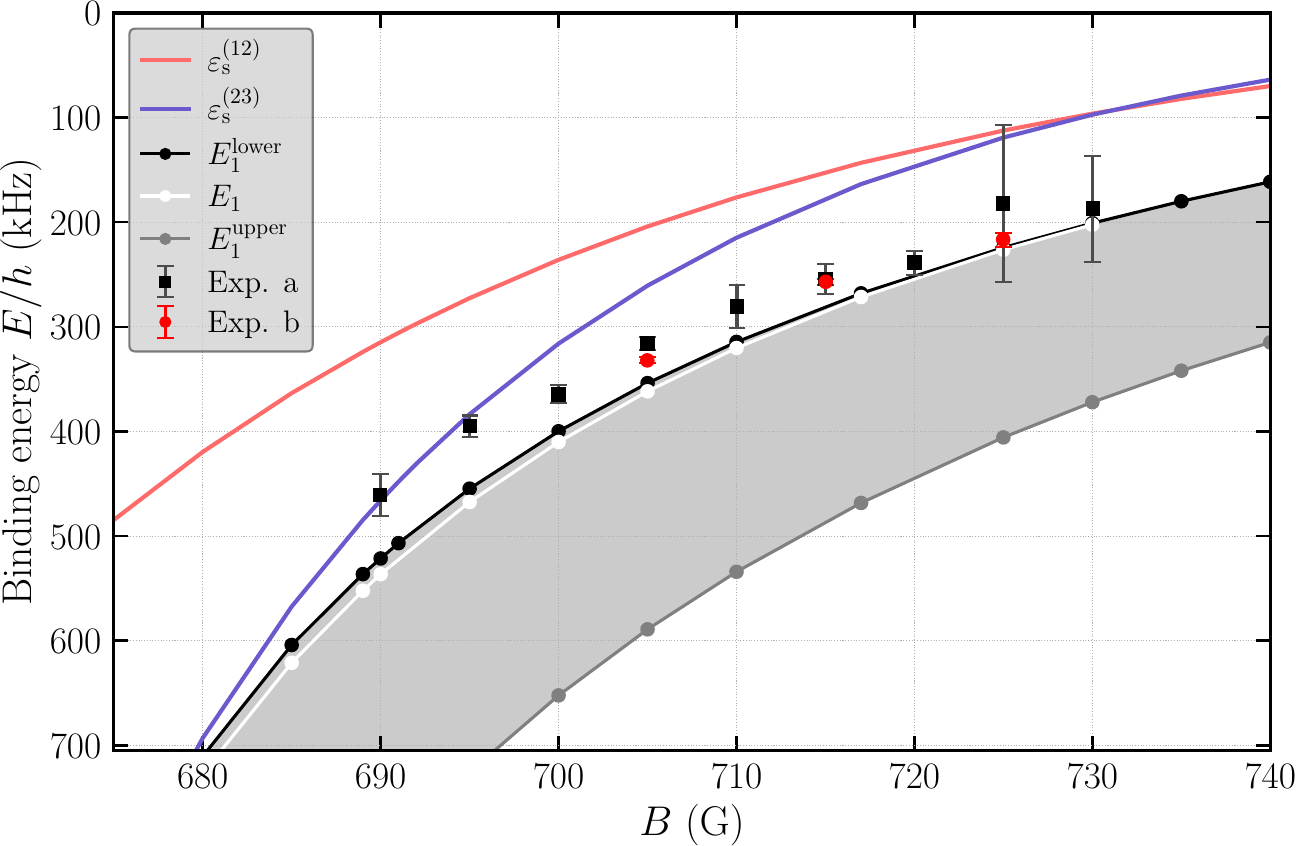}
    \caption{Calculated binding energies of the first excited Efimov trimer and two-body dimer thresholds versus $B$, alongside experimental three-body radio-frequency association data from Lompe {\it et al.} \cite{Lompe2010b}. Solid red and blue curves show the $(12)$ and $(23)$ dimer thresholds. The black, white, and gray solid curves with circular markers represent the calculated trimer binding energies obtained with the diagonal correction [\cref{eq:3BP_dia}], with three coupled channels [\cref{eq:coupledequations}], and without the diagonal correction. Black squares and red circles denote the experimental trimer binding energies extracted from Fig.~4 of Ref.~\cite{Lompe2010b} for the $\ket{1}$--$\ket{12}$ and $\ket{2}$--$\ket{23}$ initial mixtures.}
    \label{fig:3BE}
\end{figure}

The binding energy spectrum of the first excited Efimov trimer was previously measured experimentally using radio-frequency association \cite{Lompe2010b}. Our model is tested and validated through a comparison of the computed binding energies as a function of $B$ with the data of Lompe {\it et al.}, shown in \cref{fig:3BE}. The solid red and blue curves denote the $(12)$ and $(23)$ dimer binding energies, respectively, obtained by diagonalizing the two-body Hamiltonian in \cref{eq:radial schrodinger} with the $B$-dependent scattering lengths of Zürn {\it et al.} \cite{Zurn2013}. These dimer binding energies set the asymptotic limits of the effective three-body hyperradial potentials, and trimer branches can only exist below the deeper of these two curves. For $a>0$, the two-body potential [\cref{eq:twobody potential}] is tuned to support a single $s$-wave bound state, whereas for $a<0$ no bound state occurs. We compute three trimer spectra: one obtained from the three-body effective potential [\cref{eq:3BP_dia}] including the diagonal correction, which provides a rigorous lower bound on the trimer binding energies and is represented by a black solid curve with circular markers; a second obtained with the diagonal correction omitted, which gives an upper bound and is represented by a gray solid curve with circular markers; and a third obtained by solving the coupled equations in \cref{eq:coupledequations} with three coupled channels, represented by a white solid curve with circular markers. The spectra based on the effective potential with the diagonal correction (black) and on the three-channel coupled equations (white) follow the experimentally measured binding energies quite closely, both slightly overestimating the binding energies, with the agreement improving at higher $B$, where the calculated and experimental values almost coincide. In contrast, the spectrum obtained from the effective potential without the diagonal correction (gray) predicts much larger binding energies over the entire field range, showing that neglecting this correction leads to a substantial overestimation of the trimer binding energy.

\subsection{Recombination rate coefficients}
With the trimer spectrum thus established, we now turn to temperature‐dependent three‐body recombination by examining the thermally averaged rate coefficients for a nondegenerate gas at four different temperatures.

\begin{figure}[htbp!]
    \centering
    \includegraphics[width=1.\linewidth]{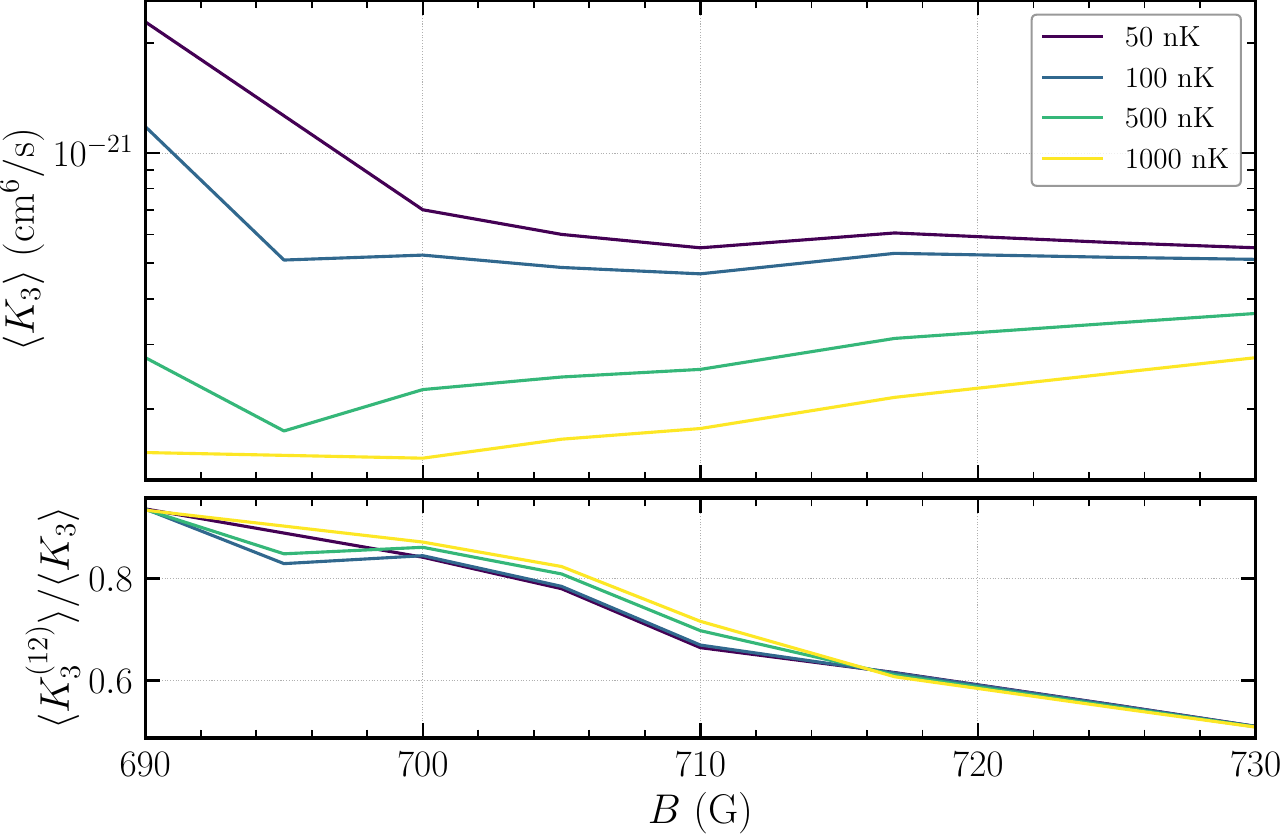}
    \caption{Thermally averaged three‐body recombination rate coefficients $\langle K_3\rangle$ (top) and branching ratios $\langle K_3^{(12)}\rangle/\langle K_3\rangle$ (bottom) as functions of $B$ at four temperatures. In the upper panel, $\langle K_3\rangle$ peaks at $690\,\rm G$ for the coldest gas and follows the expected $K_3\propto E^{-1}$ scaling near threshold as $a_{13}\to\infty$, with the inverse‐energy-scaling behavior gradually weakening as $B$ increases. In the lower panel, the branching ratio is $0.94$ at $690\,\rm G$ and decreases to $0.5$ near $730\,\rm G$.}
    \label{fig:K3_vs_B}
\end{figure}

\cref{fig:K3_vs_B} shows the recombination rate $\langle K_3\rangle$ (top) and the branching ratio $\langle K_3^{(12)}\rangle/\langle K_3\rangle$ (bottom) versus $B$ at 50, 100, 500, and 1000~nK. In the upper panel, $\langle K_3\rangle$ peaks at 690~G for the coldest gas (50~nK), and the order of the curves (50~>~100~>~500~>~1000~nK) follows the expected scaling $K_3\propto E^{-1}$ as $a_{13}\to\infty$. Moving away from 690~G, this inverse-energy-scaling behavior gradually weakens, although a clear temperature-dependent spread remains throughout the field range. In the lower panel, the branching ratio remains near 0.95 at 690~G for all temperatures, indicating an almost exclusive production of the $(12)$ dimer when $a_{13}$ diverges, and decreases to $0.5$ near 730~G, where the binding energies of the $(12)$ and $(23)$ dimers become degenerate (see \cref{fig:3BE}). The near-collapse of all curves onto a single line indicates that the branching ratio is essentially temperature-independent and that it relies on the proximity to the two dimer thresholds.  

\begin{figure}[htbp!]
    \centering
    \includegraphics[width=1.\linewidth]{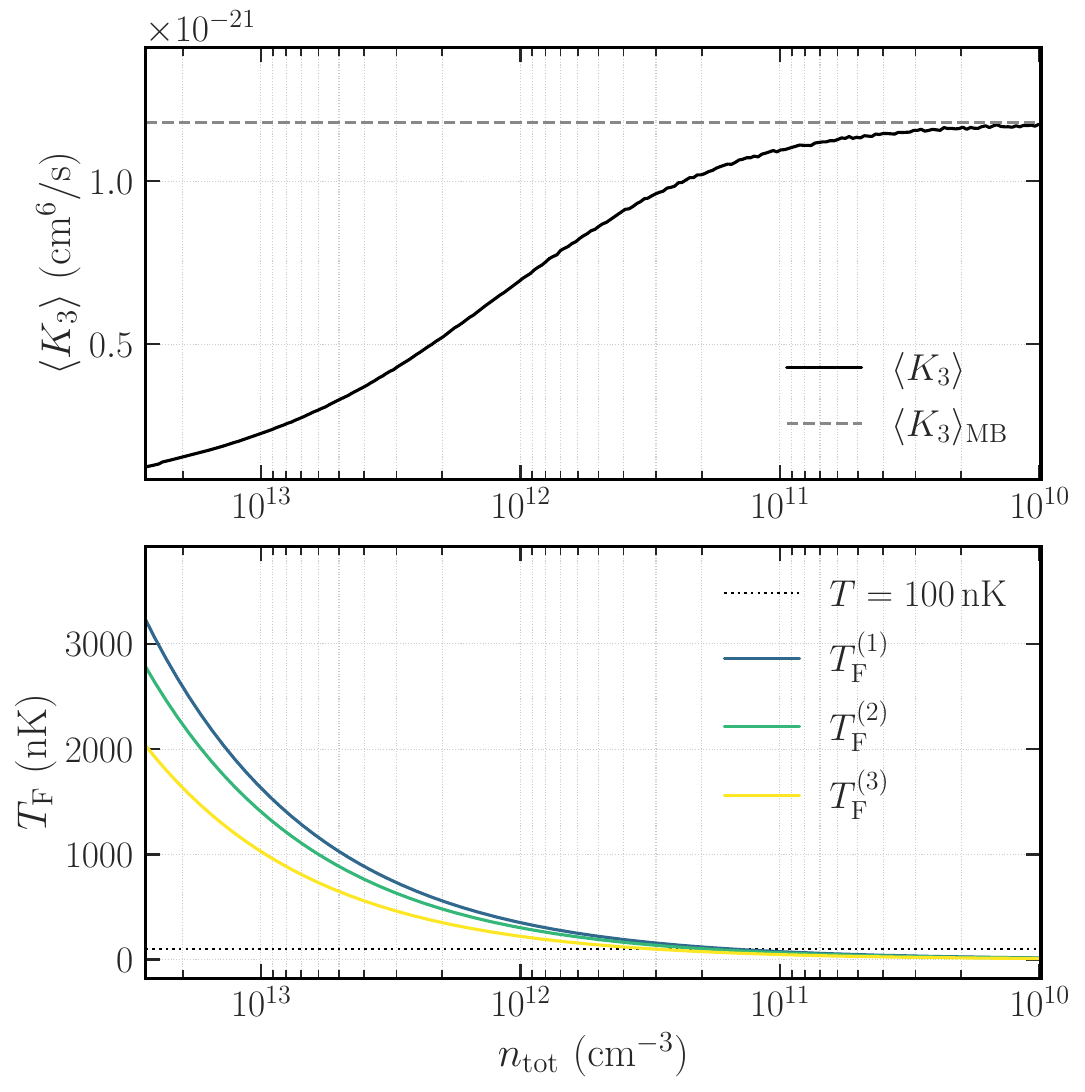}
    \caption{Upper panel: Thermally averaged three‐body recombination rate coefficient $\langle K_3\rangle$ at $B=690 \, \rm G$ as a function of total density $n_{\rm tot}$.  The solid black line is the quantum‐degenerate result, and the dashed gray line marks the classical Maxwell–Boltzmann value $\langle K_3\rangle_{\rm MB}$ at $T=100 \, \rm nK$.  Lower panel: Fermi temperatures $T_{\rm F}^{(i)}$ for each spin component (solid colored lines) and the gas temperature $T=100\, \rm nK$ (dotted line). The point where all $T_{\rm F}>T$ (around $n_{\rm tot}\sim10^{11} \, \rm cm ^{-3}$) coincides with the onset of suppression in $\langle K_3\rangle$.}
    \label{fig:K3}
\end{figure}

\cref{fig:K3} illustrates the behavior of the thermally averaged recombination rate $\langle K_3\rangle$ at $B=690\,\rm G$ as the gas enters the degenerate regime. In the upper panel, $\langle K_3\rangle$ (solid black line) is plotted against total density $n_{\rm tot}$ on a logarithmic scale, the dashed gray line showing the classical Maxwell--Boltzmann value $\langle K_3\rangle_{\rm MB}$ for $T=100\,\rm nK$. As $n_{\rm tot}$ increases (and hence $T/T_{\rm F}$ decreases), $\langle K_3\rangle$ falls below the classical prediction, indicating a suppression of three-body loss in the degenerate regime. This suppression arises because $K_3(E)\propto E^{-1}$ when one of the three scattering lengths diverges: In a degenerate gas, the mean collision energy is higher, driving the thermally averaged $\langle K_3\rangle$ downward.

The lower panel shows the Fermi temperatures $T_{\rm F}^{(i)}$ for each spin component (solid colored lines) along with the gas temperature $T=100 \, \rm nK$ (dotted line). The crossover $T_{\rm F}>T$ occurs near $n_{\rm tot}\sim10^{11} \, \rm cm^{-3}$, close to where $\langle K_3\rangle$ departs from the Maxwell--Boltzmann prediction. This explains the early-time behavior observed in~\cite{Schumacher2023}: At $B=690 \, \rm G$ and the highest initial densities, quantum degeneracy suppresses three-body loss below the classical expectation, the effect weakening as the gas dilutes. This suppression directly reflects the inverse energy scaling of $K_3(E)$ when $a_{13}\to\infty$.  

\subsection{Secondary collisions and mean free paths}
Having determined the three-body recombination rate coefficients, we now turn to the subsequent dynamics of the recombination products at $B=690\, \rm G$. \cref{tab:K3 and cross sections} lists the thermally averaged three‐body recombination coefficients together with the two‐body elastic and inelastic cross sections for collisions of the newly formed $(12)$ and $(23)$ dimers with cold atoms, as well as for collisions between a kinetic recombination atom $\ket{i}_{\rm k}$ and a cold atom. The corresponding total and spin-resolved mean free paths of the recombination products are also included. 

The elastic cross section for both atom-atom and atom-dimer collisions between two spin components is given by
\begin{equation}\label{eq:elastic 2b CS}
  \sigma^{\rm el}(k_{\rm rel}) = \frac{4\pi a(k_{\rm rel})^2}{1 + k_{\rm rel}^{2}a(k_{\rm rel})^2},
\end{equation}
where the energy‐dependent scattering length $a(k_{\rm rel})$ follows the effective-range expansion  
\begin{equation}\label{eq:energydep a}
  \frac{1}{a(k_{\rm rel})} = \frac{1}{a} - \frac{1}{2}r_{\rm e}\,k_{\rm rel}^2 + \mathcal{O}(k_{\rm rel}^4).
\end{equation}
In the regime $a\gg\bar a$, $r_{\rm e}=2.9179\,\bar a$ and the mean scattering length $\bar a = [4\pi/\Gamma(1/4)^2]\,r_{\rm vdW}$ \cite{Chin2010}.  For atom-dimer collisions, $a$ is replaced by the universal atom-dimer scattering length $a_{\rm ad}^{(ij)} = 1.18\,a_{ij}$, as given in~\cite{Giorgini2008}. 

To capture the stochastic nature of secondary collisions, a Monte Carlo sampling procedure is employed. Each recombination event releases a binding energy $\varepsilon_{\rm s}^{(ij)}$, which is partitioned so that the $(ij)$ dimer carries $\varepsilon_{\rm s}^{(ij)}/3$ and the free atom in spin $k$ carries $2\varepsilon_{\rm s}^{(ij)}/3$. Both products are assigned isotropic directions and the cold atom is assumed to either be stationary or have a velocity drawn either from a Maxwell--Boltzmann or a Fermi--Dirac distribution at $T=100\, \rm nK$. From the resulting velocities, the relative speed $v_{\rm rel}$ between the kinetic product and the cold atom is calculated. The collision energy in the center-of-mass frame is then $E_{\rm rel}=\mu_{\rm red} v_{\rm rel}^2/2$, $\mu_{\rm red}$ being the reduced mass of the collision complex, and $k_{\rm rel}$ the corresponding wave number. The energy-dependent scattering length is obtained using $k_{\rm rel}$ via \cref{eq:energydep a}, and the inelastic or elastic cross section $\sigma(k_{\rm rel})$ is evaluated from its respective defining formula [i.e., \cref{eq:cross section} and \cref{eq:elastic 2b CS}]. Finally, averaging $\sigma(k_{\rm rel})$ over many such samples yields the thermally averaged secondary‐collision cross section. \cref{tab:K3 and cross sections} lists the cross sections obtained by assuming a stationary cold atom, sampling its velocity from a Maxwell--Boltzmann distribution, and sampling from a Fermi--Dirac distribution. For the latter, equal per-spin densities $n_i = 2\times10^{11}\,\mathrm{cm}^{-3}$ are used. The same densities are also used to calculate the mean free paths. For an $(ij)$ dimer the total mean free path is
\begin{equation}
  \ell_{\rm mfp}^{(ij)} = \frac{1}{\sum_{k=1}^3 n_k \sigma_{(ij-k)}},
\end{equation}
and for a free atom in spin state $k$
\begin{equation}
  \ell_{\rm mfp}^{(k)} = \frac{1}{\sum_{i\neq k} n_i\sigma_{(k-i)}}.
\end{equation}

In the experiments of Schumacher {\it et al.}~\cite{Schumacher2023}, the atoms are confined in box traps of conical-frustum geometry. Although several sizes are used, the only explicitly specified dimensions are $L=120\,\mu\mathrm{m}$, $R_1=75\,\mu\mathrm{m}$, and $R_2=73\,\mu\mathrm{m}$, which are adopted here. At the densities of interest, whether a Maxwell--Boltzmann or Fermi--Dirac distribution is chosen for the cold-atom energies makes little difference; they both produce essentially the same thermally averaged secondary-collision cross sections (see \cref{tab:K3 and cross sections}). More striking are the mean free paths of the recombination products, which are in the $10-30\,\mu \rm m$ range. These distances are comparable to the radii of the trap and much smaller than its length, implying that most $(ij)$ dimers and energetic atoms undergo at least one secondary collision before an eventual escape. Consequently, these secondary collisions play a dominant role in both ejecting atoms from the trap and depositing energy into the remaining gas.

A typical three-body recombination event produces a shallow $(12)$ dimer and a spin-$3$ atom, both of which acquire kinetic energy and travel through the gas. Our computed cross sections show that $(12)$ dimers most often scatter off spin-3 atoms, whereas the energetic spin-$3$ atoms predominantly collide with spin-$1$ atoms. In a balanced three-spin mixture, this predicts an enhanced loss of spin-$3$ and spin-$1$ atoms compared to spin-$2$ atoms, since a single collision suffices to reduce a $(12)$ dimer's energy below the trap depth, after which it remains confined, and dimers are indistinguishable from atoms in the measurements of Ref.~\cite{Schumacher2023}. Consequently, secondary collisions alone cannot reproduce the uniform decay observed in balanced samples. To investigate this discrepancy, we have developed a Monte Carlo code that tracks the full cascade of secondary collisions and loss processes.
\begin{table}[h!]
\centering
\caption{Thermally averaged three-body recombination coefficients, secondary-collision cross sections, and mean free paths at $T = 100\,\mathrm{nK}$.  
The secondary-collision panel gives elastic cross sections for a stationary cold atom ($\sigma_0$), for Maxwell--Boltzmann-sampled ($\sigma_{\rm MB}$), and Fermi--Dirac-sampled atom energies ($\sigma_{\rm FD}$), along with the inelastic cross section for the $\ket{12}\!-\!\ket{3}$ channel.  
Mean free paths $\ell_{\rm mfp}$ for the recombination products are calculated using $\sigma_{\rm MB}$, with both species-resolved and total values presented in the lowest panel. All Fermi--Dirac cross sections and mean free path calculations assume a per-spin density of $n_i = 2\times10^{11}\,\mathrm{cm}^{-3}$.
}
\label{tab:K3 and cross sections}
\begin{tabularx}{\columnwidth}{@{}*{3}{C}N@{}}
\toprule
\multicolumn{4}{c}{\textbf{Three-body recombination}}\\
\midrule
\(\langle K_3\rangle\) (cm\(^6\)/s)
  & \(\langle K_3^{(12)}\rangle\) (cm\(^6\)/s)
  & \(\langle K_3^{(23)}\rangle\) (cm\(^6\)/s)
  & BR \\ 
\midrule
\(1.18\times10^{-21}\)
  & \(1.11\times10^{-21}\)
  & \(7.48\times10^{-23}\)
  & 0.94 \\ 
\bottomrule
\end{tabularx}
\vspace{1ex}
\begin{tabularx}{\columnwidth}{@{}l*{3}{>{\centering\arraybackslash}X}@{}}
\toprule
\multicolumn{4}{c}{\textbf{Secondary collisions}} \\ 
\midrule
Channel 
  & \(\sigma^{\rm el}_0\) (cm\(^2\))
  & \(\sigma^{\rm el}_{\rm MB}\) (cm\(^2\))
  & \(\sigma^{\rm el}_{\rm FD}\) (cm\(^2\)) \\ 
\midrule
\(\ket{12}\!-\!\ket{3}\)                     & \(3.37\times10^{-9}\) & \(3.44\times10^{-9}\) & \(3.46\times10^{-9}\) \\ 
\(\ket{23}\!-\!\ket{1}\)                     & \(2.67\times10^{-9}\) & \(2.71\times10^{-9}\) & \(2.73\times10^{-9}\) \\
\(\ket{12}\!-\!(\ket{1}\,\text{or}\,\ket{2})\) & \(8.20\times10^{-10}\) & \(8.14\times10^{-10}\) & \(8.12\times10^{-10}\) \\
\(\ket{23}\!-\!(\ket{2}\,\text{or}\,\ket{3})\) & \(5.61\times10^{-10}\) & \(5.58\times10^{-10}\) & \(5.57\times10^{-10}\) \\
\(\ket{3}_{\rm k}\!-\!\ket{1}\)             & \(2.01\times10^{-9}\) & \(2.02\times10^{-9}\) & \(2.03\times10^{-9}\) \\
\(\ket{3}_{\rm k}\!-\!\ket{2}\)             & \(3.98\times10^{-10}\) & \(3.97\times10^{-10}\) & \(3.97\times10^{-10}\) \\
\(\ket{1}_{\rm k}\!-\!\ket{2}\)             & \(4.78\times10^{-10}\) & \(4.78\times10^{-10}\) & \(4.76\times10^{-10}\) \\
\(\ket{1}_{\rm k}\!-\!\ket{3}\)             & \(1.36\times10^{-9}\) & \(1.36\times10^{-9}\) & \(1.36\times10^{-9}\) \\
\midrule
\multicolumn{4}{c}{\textbf{Inelastic \(\ket{12}-\ket{3}\) collision}} \\
\midrule
\(\ket{12}\!-\!\ket{3}\) & \(\sigma^{\rm in}_0\) (cm\(^2\)) & \(\sigma^{\rm in}_{\rm MB}\) (cm\(^2\)) & \(\sigma^{\rm in}_{\rm FD}\) (cm\(^2\)) \\ 
\midrule
  & \(2.63\times10^{-10}\) &  \(2.68\times10^{-10}\) &  \(2.70\times10^{-10}\) \\
\bottomrule
\end{tabularx}
\vspace{1ex}
\begin{tabularx}{\columnwidth}{@{}l*{4}{>{\centering\arraybackslash}X}@{}}
\toprule
\multicolumn{5}{c}{\textbf{Mean free paths} ($\mu$m)} \\ 
\midrule
Channel 
  & $\ket{1}$ 
  & $\ket{2}$
  & $\ket{3}$
  & $\ell_{\rm mfp}$
  \\ 
\midrule
\(\ket{12}\)         & 61 & 61 & 15 & 10 \\
\(\ket{23}\)         & 18 & 90 & 90 & 13 \\
\(\ket{1}_{\rm k}\)  & -- & 104 & 37 & 27 \\
\(\ket{3}_{\rm k}\)  & 25 & 126 & -- & 21 \\
\bottomrule
\end{tabularx}
\end{table}

\subsection{Monte Carlo modeling of atom loss and heating}
Using initial spin densities matching those reported in Ref.~\cite{Schumacher2023}, we simulate atom loss driven by secondary collisions under experimentally relevant conditions. The trap is modeled as a conical frustum, and recombination events are seeded uniformly within its volume. For each event, the momenta of the outgoing dimer and atom are sampled, which determines their kinetic energies and emission directions. Two inelastic channels are included: (i) $\ket{12}$--$\ket{3}$, in which a $(12)$ dimer collides with a spin-3 atom and converts into a shallow $(23)$ dimer plus a spin-1 atom (adding kinetic energy to the system without immediate loss), and (ii) $\ket{23}$--$\ket{1}$, which represents recombination into deeply bound dimers and removes all collision partners from the trap. A reliable microscopic calculation of inelastic atom-dimer recombination into deep molecular states is not available within the present model. Therefore, we use the inelastic $\ket{12}$--$\ket{3}$ cross section as a conservative upper bound for this channel. This choice reflects the expectation that deep-dimer formation is significantly less probable than the shallow-dimer inelastic process, while still allowing for such events to occur in the cascade.

As the energetic products traverse the cloud, cold-atom velocities
(Maxwell--Boltzmann or Fermi--Dirac at $T=100$\,nK) are sampled, elastic and inelastic scattering probabilities are evaluated, and any particle or dimer that escapes the trap or undergoes inelastic conversion is removed. Averaging hundreds of cascades per density setting yields converged time traces of total and spin-resolved atom losses. The evolution of deposited energy is also recorded but
is not shown.

\begin{figure*}[!t]
  \centering
  \includegraphics[width=\textwidth,height=0.85\textheight,keepaspectratio]{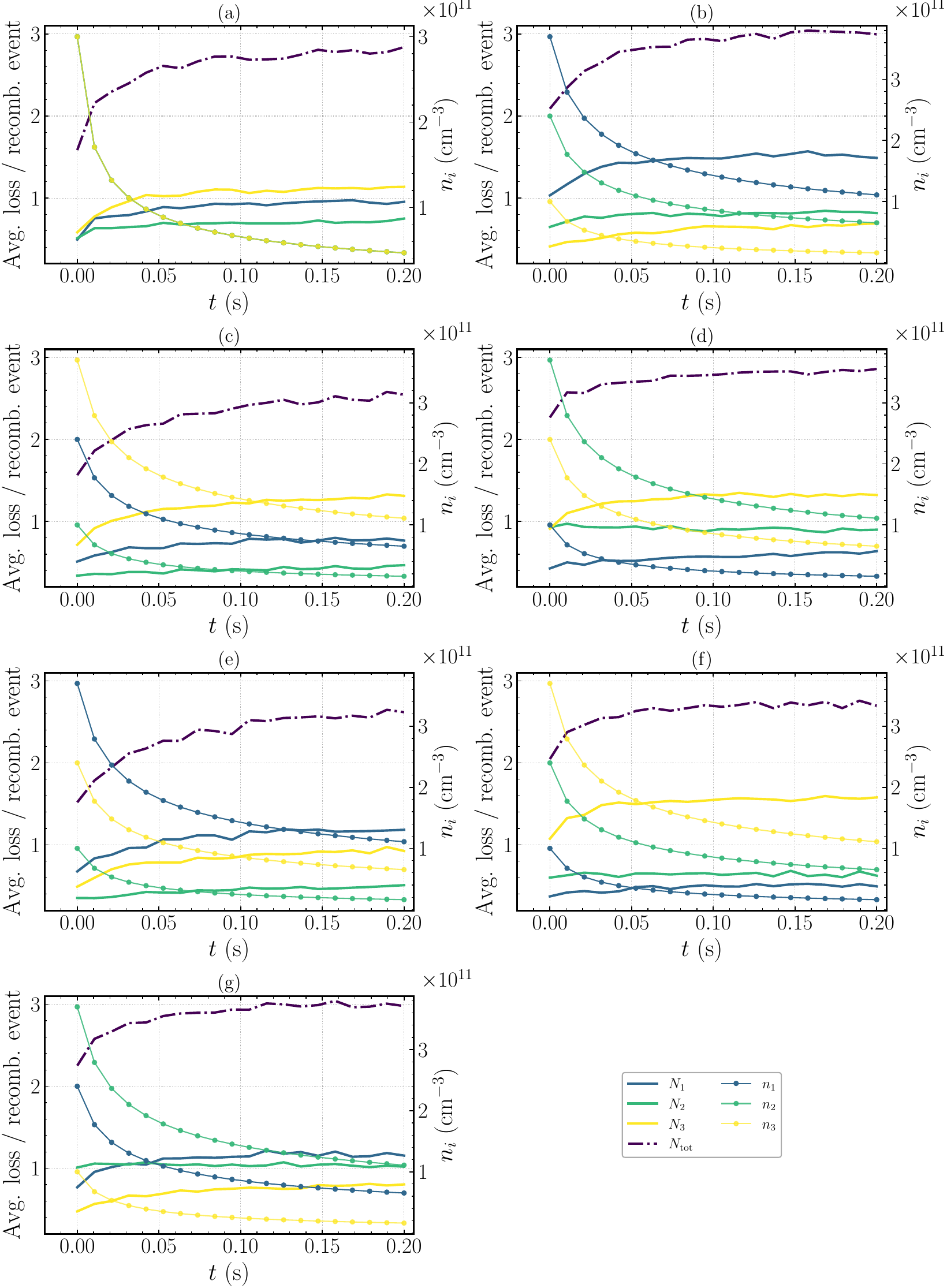}
  \caption{Average losses per recombination event versus time for experimentally relevant initial conditions. Solid lines show the spin-resolved average loss $N_{i}(t)$ from Monte Carlo simulations; the dash-dot curve is $N_{\mathrm{tot}}(t)$. Circle markers indicate reference density trajectories $n_{i}(t)$ (right axes). For the balanced case in panel~(a), the circles are obtained by propagating the standard three-body loss model  using balanced initial spin populations consistent with the atom numbers reported in Ref.~\cite{Schumacher2023}. For the imbalanced cases in panels~(b)--(g), the circles show trajectories generated by propagating the phenomenological three-body three-spin model of Ref.~\cite{Schumacher2023}, initialized with the same reported initial spin densities (and their permutations for the six possible variants).}
  \label{fig:MC}
\end{figure*}

The Monte Carlo cascades are performed on top of a prescribed background density evolution $n_i(t)$, which enters directly into the simulation and is shown explicitly in \cref{fig:MC} as circle markers. For the balanced case, the time-dependent densities are obtained by propagating the standard three-body loss equation [\cref{eq:standard_rate_eq}] using balanced initial spin densities
consistent with the atom numbers reported in Ref.~\cite{Schumacher2023}. For the imbalanced cases, the density trajectories are generated by propagating the phenomenological spin-dependent rate-equation model introduced in Ref.~\cite{Schumacher2023} [\cref{eq:phenomenological_rate_eq}] using the same reported initial imbalanced spin densities (with all permutations across the three
spin states to explore the six possible variants). Thus, the $n_i(t)$ curves plotted as circle markers in \cref{fig:MC} serve simultaneously as the density backgrounds used in the Monte Carlo simulations and as reference trajectories for visual comparison with the resulting spin-resolved losses.

Panel~(a) of \cref{fig:MC} (balanced gas) shows spin~3 depleting fastest, followed by spin~1. This ordering reflects the dominant microscopic pathways: Recombination predominantly forms $(12)$ dimers, which most frequently collide with spin-3 atoms, causing the scattered spin-3 atom to leave the trap while the dimer typically remains confined. Likewise, the energetic spin-3 atom produced in the initial recombination event either escapes directly or transfers momentum to a
spin-1 atom that may subsequently leave the trap.

Panels~(b)--(g) of \cref{fig:MC} show the six imbalanced configurations (with distinct labelings of the same initial density triplet). Across both balanced and imbalanced cases, the total average loss per event typically remains below three atoms over the density range shown, indicating that a portion of the released binding energy is retained in the gas (net heating). In the present simulations,
the temperature is held fixed for all densities. Allowing $T$ to evolve would introduce a closed feedback loop: Changes in temperature modify the cold-atom momentum distribution, which affects the recombination rates and scattering cross sections, and therefore feeds back into the density evolution itself. Incorporating this dynamical coupling is beyond the scope of the present model but is not required to capture the qualitative trends discussed here.

In general, the per-spin loss rate increases with the local density of that spin [see panels~(b), (c), (e), and (f) of \cref{fig:MC}], consistent with the phenomenological decay law of Ref.~\cite{Schumacher2023}. However, density alone does not determine the loss ordering: Channel-dependent scattering cross sections
bias loss toward spins 3 and~1 relative to spin~2 [see panels~(d) and~(g) of \cref{fig:MC}]. \textit{This channel-induced bias contrasts with the nearly identical loss rates observed experimentally for the three spin components in the balanced sample of Ref.~\cite{Schumacher2023}.}

Simulations repeated for a substantially smaller trap (cone with $L = 30\,\mu$m and $R_1 = R_2 = 15\,\mu$m) show the same early-time behavior: The average loss initially remains below three (heating). At later times, as the densities decrease, the average loss approaches three or slightly exceeds it, corresponding to a small negative energy balance (net cooling), since more kinetic energy is carried out of the system than is deposited. Taken together, these trends indicate net
energy deposition (heating) under typical conditions, with departures toward slight cooling only in the smallest trap at the lowest densities.

The limited average loss arises because a newly formed dimer typically undergoes only one or two secondary collisions before its kinetic energy falls below the escape threshold, after which it remains confined. Thus, fewer than three atoms escape per recombination event on average, and not all of the released binding energy leaves the trap. The remaining energy is retained as kinetic energy of trapped dimers and atoms, heating the gas and driving evaporative loss.
Consequently, the enhanced losses observed in Ref.~\cite{Schumacher2023} could stem from a combination of secondary (and higher-order) collision-induced ejecta and evaporation over the trap barrier.

It is also tempting to speculate that, because the different spin-pair channels in our three-component gas occupy distinct interaction regimes, energy deposited by secondary collisions will heat them at different rates, potentially giving rise to spin-dependent temperature imbalances. In particular, the 1--3 pair resides very close to the unitary limit, where two-component Fermi gases are known to exhibit an anomalously large specific heat capacity $C_V$ near the superfluid transition at degeneracies close to that used in the experiment \cite{Wyk2016,Ku2012}. Even a modest fraction of the recombination binding energy deposited into that channel will therefore produce only a small temperature increase, whereas the 1--2 and 2--3 pairs, lying further from unitarity, have more normal heat capacities and would heat more readily under the same energy input. As a result, secondary collisions could not only drive overall heating, but also amplify temperature differences between spin components, a mechanism that could help explain why samples with a majority of spin-2 atoms in~\cite{Schumacher2023} exhibited higher apparent loss rates.

\section{Conclusions}
We have solved the three-body problem for three distinguishable spin states in a Fermi gas using the hyperspherical adiabatic representation together with a single-channel, van der Waals two-body interaction model. Employing the eigenchannel $R$-matrix method, we have computed the $S$-matrix and extracted both recombination rate coefficients and two-body elastic cross sections. After thermally averaging the recombination rate coefficients in the degenerate regime, we incorporated the channel-dependent branching ratios and elastic cross sections into a Monte Carlo cascade simulation of secondary collisions and trap escape. This simulation shows that each recombination event removes fewer than three atoms on average, indicating that a portion of the released binding energy remains trapped as kinetic energy.

Based on these findings, we hypothesize that the atom loss observed in~\cite{Schumacher2023} arises from two intertwined mechanisms: (i) the ejection of secondary collision products following three-body recombination, and (ii) the evaporative loss driven by the retained kinetic energy, which heats the trapped gas. Crucially, because the interaction of the 1--3 spin pair is very close to unitarity and the experiment is conducted at temperatures just above, but not far from, the superfluid phase transition temperature $T_{\rm c}$, its heat capacity is significantly enhanced relative to the 1--2 and 2--3 pairs. We therefore speculate that the different spin components heat at distinct rates under these conditions.

A fully quantitative description of these coupled loss-heating dynamics remains a substantial theoretical challenge for several reasons. Firstly, the heat capacity of a three-component Fermi gas with unequal interaction strengths is not well known, especially near $T_{\rm c}$, where pairing correlations and strong interactions can dramatically alter thermodynamic properties. Secondly, the three-body recombination rate coefficients depend on both temperature and density, and must be computed through thermal averaging. Thirdly, the evaporation and heating rates depend on both the temperature and density of the gas, which in turn evolve due to loss and rethermalization processes. These interdependencies render the problem highly nonlinear and time dependent, with feedback between loss, heating, and evaporation that is difficult to disentangle. Capturing these effects will require self-consistent models that combine microscopic recombination physics with finite-temperature many-body thermodynamics and kinetic theory.

\section*{Acknowledgements}

We gratefully acknowledge discussions with and access to unpublished data from Nir Navon and Grant Schumacher.  This work was supported in part by NSF grants PHY-2207977 and PHY-2512984.

\clearpage

\bibliography{bibliography}

\end{document}